\documentclass[11pt]{article}
\usepackage{amsfonts,amsmath,amssymb}
\usepackage{enumerate}
\usepackage{hyperref}
\usepackage{bbm}
\usepackage{nicefrac}
\usepackage[all]{xy}
\usepackage{graphicx}

\usepackage{booktabs}
\newcommand{\ra}[1]{\renewcommand{\arraystretch}{#1}}

\newcommand{\be}{\begin{equation}}
\newcommand{\ee}{\end{equation}}

\newcommand{\bea}{\begin{eqnarray}}
\newcommand{\eea}{\end{eqnarray}}

\newcommand{\bes}{\begin{subequations}}
\newcommand{\ees}{\end{subequations}}

\newcommand{\cN}{{\cal N}}
\newcommand{\cW}{{\cal W}}

\usepackage{multirow}
\usepackage{rotating}

\begin{document}

\makeatletter
\renewcommand{\theequation}{\thesection.\arabic{equation}}
\@addtoreset{equation}{section}
\makeatother

\begin{titlepage}

\begin{flushright}
ICCUB-13-233 \\
CPHT-RR051.1015 
\end{flushright}

\vspace{25pt}

\begin{center}
   \baselineskip=16pt
   \begin{Large}\textbf{
Electric/magnetic duality \\[12pt]  and RG flows in AdS$_4$/CFT$_3$}
   \end{Large}
   		
\vspace{25pt}
		
{Javier Tarr\'io$^{1}$ and Oscar Varela$^{2,3}$}
		
\vspace{25pt}

	\begin{small}

	{\it ${}^1$ Departament de F\'isica Fonamental and Institut de Ci\`encies del Cosmos,\\
	Universitat de Barcelona, Mart\'\i \ i Franqu\`es 1, 08028 Barcelona, Spain } \\
	j.tarrio@ub.edu

	\vspace{15pt}
		
	{\it ${}^2$	Center for the Fundamental Laws of Nature,\\
	Harvard University, Cambridge, MA 02138, USA } \\
	ovarela@physics.harvard.edu

	\vspace{15pt}
		
	{\it ${}^3$	Centre de Physique Th\'eorique, Ecole Polytechnique, CNRS UMR 7644 \\
	91128 Palaiseau Cedex, France
	 } \\
		
	\end{small}

\vskip 50pt

\end{center}

\begin{center}
\textbf{Abstract}
\end{center}

\begin{quote}

A large set of relevant deformations of the ABJM field theory defined on a stack of M2 branes is captured holographically by $D=4$ $\cN=8$ SO(8)-gauged supergravity, which has accordingly been applied to study renormalisation group (RG) flows of the field theory between distinct superconformal phases. Recently, it has been discovered that this supergravity is not unique, and that a one-parameter family of gaugings of maximal supergravity exists. The parameter is an angle that measures the mixture of electric and magnetic vectors that gauge SO(8) dyonically. We construct and comprehensively characterise all domain walls between the critical points of the new supergravities with at least SU(3) invariance, which are expected to be dual to RG flows of new field theories defined at least at large $N$. We also construct some walls running off to infinity in scalar space, which we expect to be dual to Coulomb branch flows of these field theories.

\end{quote}

\vfill

\end{titlepage}

\tableofcontents


\section{Introduction}

The three-dimensional superconformal field theory defined on $N$  coincident M2-branes probing an orbifold singularity $\mathbb{C}^4/\mathbb{Z}_k$ has been proposed by ABJM \cite{Aharony:2008ug}, building on previous work \cite{Bagger:2006sk, Gustavsson:2007vu}, to comprise two copies of Chern-Simons theory at levels $k$, $-k$, with gauge group U$(N) \times$U$(N)$, and coupled to bifundamental matter with interactions dictated by a quartic superpotential. For $k >2$, this theory has manifest $\cN =6$ supersymmetry and SO(6) R-symmetry, and becomes weakly coupled at large $k$. When $k=1$ or $k=2$, on the other hand, the theory is strongly coupled and, although not manifest in the formalism of \cite{Aharony:2008ug}, it has been argued to have its supersymmetry and  R-symmetry enhanced to $\cN =8$ and SO(8), respectively. From an eleven-dimensional perspective, the near-horizon geometry of the $k=1$ stack of $N$ M2-branes corresponds to the maximally supersymmetric  Freund-Rubin background AdS$_4 \times S^7$.

Large classes of well defined AdS$_4$/CFT$_3$ dual pairs are now known with further reduced supersymmetry. For example, shortly after the ABJM theory was proposed, an $\cN=2$ superconformal U$(N) \times$U$(N)$ Chern-Simons theory, with U(1) R-symmetry and SU(3) global symmetry, coupled to matter with interactions described by a sextic superpotential, was constructed in \cite{Benna:2008zy}. This theory was further conjectured in that reference to be dual to M-theory on an $\cN=2$ AdS$_4 \times S^7$ background constructed earlier in \cite{Corrado:2001nv}, where now the product is warped, and the metric on $S^7$ is stretched and squashed so that its isometry is SU(3)$\times$U(1)$^2$ (a symmetry which is reduced to SU(3)$\times$U(1), as in the dual field theory, by the presence of four-form internal fluxes). Further support for this conjecture was given in \cite{Klebanov:2008vq}, where the SU(3)$\times$U(1) quantum numbers and conformal dimensions of the Kaluza-Klein spectrum of M-theory on the AdS$_4$ solution of  \cite{Corrado:2001nv} was argued to match the spectrum of certain operators in the conjectured dual field theory.

Having M-theory duals involving (different) metrics and flux configurations on (the same) internal space, $S^7$, both field theories \cite{Aharony:2008ug} and \cite{Benna:2008zy} are in fact related: as discussed in \cite{Benna:2008zy}, the latter arises as the infrared (IR) fixed point of the renormalisation group (RG) flow triggered by certain supersymmetric mass deformation of ABJM. As a matter of fact, the (large $N$) flow itself had been constructed holographically much earlier \cite{Ahn:2000aq,Ahn:2000mf}, within the formalism of domain walls of gauged supergravity. In this respect, the SO(8)-gauged, $\cN =8$ supergravity \cite{deWit1} of de Wit and Nicolai proves to be an extremely helpful venue to holographically study certain aspects of  strongly-coupled, low level ABJM (in the large $N$ regime), due to its origin  as a consistent truncation of $D=11$ supergravity on $S^7$ \cite{deWit2,Nicolai:2011cy,deWit:2013ija}. Particularly, a large class of conformal phases of large $N$, low level ABJM and the possible RG connections between them can be economically studied from the gauged supergravity.

Firstly, by virtue of the consistency of the truncation, all solutions of the SO(8)-gauged, $\cN =8$ supergravity \cite{deWit1} give rise to well defined M-theory backgrounds upon uplift on $S^7$. From a holographic perspective, this ultimately guarantees that the large $N$ field theories studied from the supergravity do have a well defined quantum mechanical extension at finite $N$. Secondly, the fields of the gauged supergravity correspond to all the lowest Kaluza-Klein modes of M-theory on AdS$_4\times S^7$. In particular, the supergravity scalars are formally dual to all possible mass deformations for the bifundamentals of ABJM. Finally, the (supersymmetric) AdS critical points of the gauged supergravity correspond to (super)conformal phases of the field theory. The $\cN=8$ SO(8)-gauged supergravity  \cite{deWit1} can thus be used to holographically study all possible conformal phases of low level ABJM (at large $N$) that arise as IR fixed points of RG flows triggered by mass deformations of the theory in its maximally supersymmetric, SO(8)-symmetric conformal phase in the ultraviolet (UV). Although this does not exhaust {\it all} conformal phases or RG flows between them in the multiple M2 brane field theory, it does provide a holographic account of a large and interesting class of them\footnote{Indeed, deformations of the SO(8) conformal phase exist which, although still relevant, do not correspond to mass terms \cite{F-theorem,Gabella:2012rc}. These are thus left outside the truncation to $D=4$ $\cN=8$ supergravity, and are not necessarily described holographically by a consistently truncated $D=4$ supergravity. The supergravity dual of the IR conformal phase corresponding to a deformation of this type has been constructed, directly in eleven-dimensions, in \cite{Gabella:2012rc,Halmagyi:2012ic}. Another type of mass-driven RG flows that cannot be described within the $\cN=8$ theory are those for which the 'skew-whiffed' SO(8) phase of the anti-M2 brane field theory arises in the IR, rather than in the UV. Some of these flows can be studied holographically using the consistent truncations of \cite{Cassani:2011fu,Cassani:2012pj}.}.

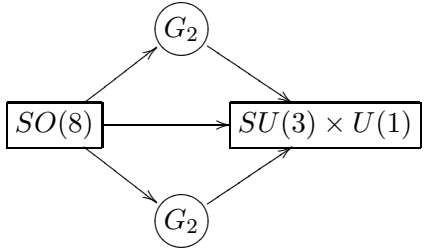
\begin{figure}[tb]
\[
\xymatrix@C=0.2cm@R=0.2cm{
 && *++[o][F]{G_2}\ar@{->}[rdd] &  \\
 &&&\\
 *+[F]{SO(8)}  \ar@{->}[rrdd] \ar@{->}[rruu] \ar@{->}[rrr] 
 & & & *+[F]{SU(3)\times U(1)} \\
 &&&\\
 & & *++[o][F]{G_2}\ar@{->}[ruu]  &
 }
\]
\caption{Schematic view of the ($\omega =0$) supersymmetric critical points with at least SU(3) invariance and the supersymmetric flows connecting them represented by arrows pointing from the UV to the IR. The G$_2$ points and the flows with them as endpoints are physically equivalent (they are doubled here in order to reflect a redundancy in the parametrisation that we use in section \ref{sec.flows}). Further, a one-parameter family of flows (or cone of flows) exists between SO(8) and SU(3)$\times$U(1).}\label{fig.BHPW}
\end{figure}

The above examples all fit into this scheme. Indeed, the scalar potential of the $\cN=8$ SO(8)-gauged supergravity of \cite{deWit1} has a critical point that preserves the full $\cN=8$ supersymmetry and SO(8) symmetry of the supergravity, and another point that spontaneously breaks them down to $\cN=2$ and SU(3)$\times$U(1) \cite{Warner:1983vz}. These uplift on the $S^7$ to the well-defined Freund-Rubin and Corrado-Pilch-Warner \cite{Corrado:2001nv} backgrounds, respectively, wich are in turn dual to the M2 brane field theory in its ABJM, $\cN=8$, SO(8)-symmetric phase  \cite{Aharony:2008ug} and in the $\cN=2$ SU(3)$\times$U(1)-symmetric phase of \cite{Benna:2008zy}. The RG flow connecting the former in the UV to the latter in the IR was addressed holographically in \cite{Ahn:2000aq} within the $D=4$ gauged supergravity, and uplifted to $D=11$ in \cite{Corrado:2001nv}, providing an AdS$_4$/CFT$_3$ counterpart of similar constructions in $D=5$ and type IIB \cite{Freedman:1999gp,Pilch:2000fu}. Further checks, matching the free energies of the dual three-dimensional field theories and the volumes of the internal $S^7$ have been performed in \cite{F-theorem, Gabella:2011sg}, thus completing a beautifully consistent holographic picture. By now, all the supersymmetric RG flows interpolating between superconformal phases of the M2 brane field theory with at least SU(3) global invariance, have been constructed \cite{Ahn:2000aq,Ahn:2001kw,Ahn:2008ya,Ahn:2008gda,Bobev:2009ms} as domain walls of the  $\cN=8$ SO(8)-gauged supergravity \cite{deWit1} between all possible critical points within the SU(3)-invariant sector \cite{Warner:1983vz} of the supergravity. Only three supersymmetric critical points exist in this sector, with SO(8), G$_2$, and SU(3)$\times$U(1) bosonic symmetry, and $\cN=8$, $\cN=1$ and $\cN=2$ supersymmetry. A continuous family of supersymmetric flows, or  {\it cone of flows}, exists \cite{Bobev:2009ms} interpolating between the $\cN=8$ and $\cN=2$ points. The family includes the direct flow of  \cite{Ahn:2000aq}, and is bounded by the latter and by flows that end or start at the G$_2$ point. See figure \ref{fig.BHPW} for a sketch of this situation\footnote{Some non-supersymmetric flows have been built in this sector \cite{Gauntlett:2009bh}, too. These were originally constructed in a subsector of the $\cN=2$ gauged supergravity that arises from consistent truncation of M-theory on an arbitrary Sasaki-Einstein sevenfold \cite{Gauntlett:2009zw}, later realised \cite{Bobev:2010ib} to match the SU(4)$\supset$SU(3) invariant sector of $\cN=8$ SO(8)-gauged supergravity \cite{deWit1}.}.

Consistently truncated supergravity is very helpful to holographically study not only the RG evolution of a field theory triggered by relevant deformations, but also that caused by vacuum expectation values (vevs) leading the theory away from a given superconformal phase. The typical bulk counterpart of this type of RG flows is again provided by domain walls,  now supported by supergravity scalars that usually run off to infinite values of the scalar potential. Interestingly, although such solutions accordingly develop a seemingly pathological singularity, the latter is usually resolved when the solution is uplifted into its parent higher-dimensional supergravity. This is made possible by the warp factors that are usually involved in consistent truncation formulae such as those of \cite{deWit2}: the asymptotics of the warping usually corrects the singular behaviour, rendering the uplifted, higher-dimensional solution singularity-free. Following and generalising the analog type IIB construction of \cite{Freedman:1999gk}, this type of domain walls in $D=4$ $\cN=8$ SO(8)-gauged supergravity \cite{deWit1} were constructed in \cite{Cvetic:1999xx}. The smooth, uplifted solutions were shown to correspond to M2 branes continuously distributed, rather than stacked, thereby describing the dual field theory in a spontaneous gauge symmetry breaking, or Coulomb, phase. This interpretation from an eleven-dimensional perspective thus lends strong support to the argument that the corresponding $D=4$ domain walls do describe holographically RG evolution triggered by vevs in the field theory.

This closed and coherent picture was somewhat shaken recently when it was pointed out \cite{Dall'Agata:2012bb} that the de Wit and Nicolai supergravity \cite{deWit1} is not unique, but rather is only a member of a one parameter family of SO(8) gaugings of maximal four-dimensional supergravity. Each supergravity in the family has local $\cN=8$ supersymmetry and SO(8) gauge symmetry, with the same embedding of SO(8) into the global symmetry group E$_{7(7)}$ of the ungauged theory. The new feature is that SO(8) is gauged dyonically \cite{Dall'Agata:2012bb}, with respect to the electric frame of the original theory of \cite{deWit1}. Schematically, the parameter characterising the family is an angle, $\omega$, that measures the linear combination $g \cos \omega \ A_{\mu}  + g \sin \omega \ \tilde A_{\mu}$ of electric, $A_{\mu}$, and magnetic, $\tilde A_{\mu}$, gauge fields, in the adjoint of SO(8), that participate in the gauging. In the ungauged limit, $g\rightarrow 0$, the rotation by $\omega$ can always be undone by an  E$_{7(7)}$ electric/magnetic duality transformation, as all electric/magnetic duality frames become equivalent in this limit. At finite coupling constant $g$, however, electric/magnetic duality is broken and the theory is typically sensitive to the duality frame chosen to introduce the gauging\footnote{ \label{fnt:rangeomega} Members of the family with $\omega \in (\frac{\pi}{8} , 2\pi)$ are, however, related by field redefinitions to members with $\omega \in [0, \frac{\pi}{8}]$ \cite{Dall'Agata:2012bb} (see also \cite{deWit:2013ija}). Distinct theories, unrelated by local field redefinitions, are thus only obtained for the latter range of $\omega$. The original theory of \cite{deWit1} is the $\omega=0$ member of the family.}. An $\omega$ dependence thus shows up in the physical couplings of the gauged supergravity and, particularly, in the scalar potential.

All critical points of the $\omega = 0$ theory \cite{deWit1} remain critical points of the family of \cite{Dall'Agata:2012bb}, although both their location in the E$_{7(7)}/$SU(8) scalar manifold and their associated cosmological constant (the value of the potential at the critical point) typically depend on $\omega$. Furthermore, new AdS critical points with no $\omega = 0$ counterpart arise. These new points either partner with other points which do remain $\omega =0$ critical points (in the sense that they share the same symmetries and mass spectra) or are altogether new in the $\omega \neq 0$ theories and have no counterpart whatsoever in the $\omega = 0$ theory. Intriguingly enough, all known supersymmetric points and many non-supersymmetric ones display $\omega$-independent spectra, although non-supersymmetric points are known which exhibit masses that do run with $\omega$. All critical points of the $\omega$-dependent family of $\cN=8$ SO(8) gaugings have now been classified in the G$_2$ \cite{Dall'Agata:2012bb,Borghese:2012qm}, SU(3) \cite{Borghese:2012zs} and SO(4) \cite{Borghese:2013dja} invariant sectors. In the SU(3) sector, for example, new supersymmetric SU(3)$\times$U(1) and G$_2$ points arise which partner with their $\omega =0$ counterparts, and an altogether new supersymmetric SU(3) point exists with no $\omega = 0$ equivalent\footnote{We will focus here on the family of SO(8) gaugings, whose only known critical points are AdS. The same dyonic construction can be performed for other gauge groups that lead to scalar potentials with de Sitter or Minkowski vacua, see \cite{DallAgata:2011aa, Kodama:2012hu,Dall'Agata:2012sx,Blaback:2013sda,Catino:2013ppa}.}.

This wealth of new SO(8) gaugings of maximal supergravity prompts an obvious question: just like the $\omega=0$ supergravity captures holographically all possible mass terms for ABJM, as discussed above, can a similar AdS$_4$/CFT$_3$ correspondence be established for each of the new, $\omega \neq 0$, gaugings? The three-dimensional CFTs should be related to the ABJM theory although, as suggested in \cite{Dall'Agata:2012bb} itself, the full quantum mechanical correspondence may lead to a discretisation of $\omega$. The existence of such CFT$_3$ dual pairs with a well defined extension of the field theory or field theories into finite $N$ is crucially related to the existence of an M-theory origin for the new gaugings, similarly to the $\omega = 0$ case. The latter issue was investigated in \cite{deWit:2013ija} (see also \cite{Godazgar:2013dma,Godazgar:2013pfa}) but, although some conditions were pinned down on the generic features that the formalism of \cite{deWit2} would impose on a possible $D=11$ embedding of the new gauged supergravities, no conclusive answer was provided and the question was left open. Other comments on the possible higher-dimensional origin of the new gaugings also appear in \cite{Aldazabal:2013mya}.

Even if the string or M-theory origin of the new gaugings \cite{Dall'Agata:2012bb} remains unclear at this stage, exploring the features of their dual field theories is nevertheless of unquestionable interest. These new supergravities correspond to (or rather, define) dual field theories in the strict large $N$ limit, about which abundant information can be inferred. In this paper, we numerically construct and exhaustively classify all possible supersymmetric domain walls between supersymmetric critical points of the new theories \cite{Dall'Agata:2012bb} with at least SU(3) invariance, that were classified in \cite{Borghese:2012zs}\footnote{Just prior to submission of this paper, some of these domain walls have also been constructed in \cite{Guarino:2013gsa}.}. We also initiate the study of a class of domain walls that interpolate between the $\cN=8$-supersymmetric, SO(8)-symmetric point that all the supergravities in the family  \cite{Dall'Agata:2012bb} display, and run off to infinite values of the supergravity fields. Both types of domain walls should holographically describe the behaviour under renormalisation group of the corresponding family of large $N$ field theories: the former, their RG evolution under deformations by certain relevant operators; the latter, should be dual to the flows originated by vevs that place the field theories in a Coulomb branch. Although we will not be able to establish precise  relations between fully-fledged quantum field theories like those of \cite{Aharony:2008ug,Bagger:2006sk,Gustavsson:2007vu} and \cite{Benna:2008zy}, to $D=4$ bulk  solutions like that in \cite{Ahn:2000aq}, and their M-theory uplifts, like \cite{Corrado:2001nv}, we will uncover interesting new behaviour of the family of large $N$  field theories not shared by the standard M2-brane field theory \cite{Aharony:2008ug}. 

\begin{figure}[tb]
\[
\xymatrix@C=0.2cm@R=0.5cm{
&  & & *++[o][F]{G_2} \ar@/^/[dr] \ar@/_/[lld]  &   \\
& *+[F]{SU(3)}  & & & *+[F]{SU(3)\times U(1)}  \\
& & *+[F]{SO(8)}  \ar[dr] \ar[rru] \ar[ruu] \ar[ul] \ar[lld]  \ar[ldd]  & &  \\ 
 *++[o][F]{G_2}  \ar@/^/[uur] \ar@/_/[dr] & && *++[o][F]{\overline{G_2}}  \ar@/^/[lld] \ar@/_/[uur] &\\
& *+[F]{SU(3)\times U(1)} & &  & 
}
\]
\caption{ Schematic view of the $\omega =\frac{\pi}{8}$ supersymmetric critical points with at least SU(3) invariance and the supersymmetric flows connecting them represented by arrows pointing from the UV to the IR. Two of the G$_2$ points and the flows with them as endpoints are physically equivalent, and the same holds for the SU(3)$\times$U(1) points and flows (they are doubled here in order to reflect a redundancy in our parametrisation). There also exist cones of flows which are dominated by the SU(3) and SU(3)$\times$U(1) points in the IR.}\label{fig.FPs}
\end{figure}
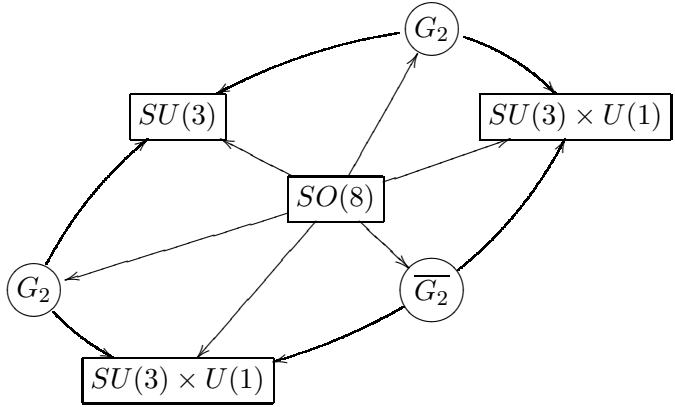

In particular, we discover a rich pattern of flows between supersymmetric critical points in the SU(3)-invariant sector. We recover the $\omega =0$ analysis of \cite{Bobev:2009ms}, and find it qualitatively different from two other cases. These arise for generic values of the dyonically gauging angle $\omega$ strictly inside its allowed range, $0 < \omega < \frac{\pi}{8}$, and when the angle attains its rightmost extremum, $\omega = \frac{\pi}{8}$. The central, maximally supersymmetric SO(8) point remains, for all $\omega$, the UV origin of a web of flows with IR end in each of the remaining points. The G$_2$ points can serve both as the IR end of flows with SO(8) UV origin, and as the UV origin of flows with IR end in the SU(3) and SU(3)$\times$U(1) points. In addition, the latter dominate the IR physics of new one-parameter families, or cones, of flows with UV origin in the SO(8) point. The cones include a distinct {\it direct} flow, in the sense that minimises the path in field space, and other flows that can approach the limiting G$_2$ points without ever reaching them. It is also very interesting to observe that, like in the $\omega = 0$ case, some identifications must be performed on physical grounds between the critical points and flows at $\omega = \frac{\pi}{8}$. These identifications add further richness into the pattern of flows and govern whether a cone of flows is bounded by flows with origin or end in a G$_2$ point only, or also by the respective direct flows. Figure \ref{fig.FPs} graphically sketches the situation for $\omega =\frac{\pi}{8}$, and table \ref{Table:cones} summarises the possible cones of flows that exist for each value or range of $\omega$. 

We make all these relevant deformation flows explicit in section \ref{sec.flows}, while we deal with the Coulomb branch flows in section \ref{sec.coulombbranch}, before concluding in section \ref{sec.conclusions}. Appendix \ref{sec:Wilson} contains a holographic analysis, associated to the Coulomb flows, of Wilson loops. We believe the latter to be new in spite of focusing on $\omega =0$.

\begin{table*}\centering
\ra{1}
{\small
\begin{tabular}{@{}rcccccccc@{}}\toprule
& \multicolumn{2}{c}{$\omega = 0$} & \phantom{abc}& \multicolumn{2}{c}{$0 < \omega < \frac{\pi}{8}$} &
\phantom{abc} & \multicolumn{2}{c}{$\omega = \frac{\pi}{8}$}\\ \cmidrule{2-3} \cmidrule{5-6} \cmidrule{8-9}
&   $\#$  & boundary && $\#$ & boundary && $\#$  & boundary \\ \midrule
$\cN=2$, SU(3)$\times$U(1)  
       & 1 & G$_2$ and direct  &&  2  &  G$_2$  && 1  & G$_2$ \\[14pt]
$\cN=1$, SU(3)
       &  0  &  $-$                 &&  1  &     G$_2$  && 1  & G$_2$ and direct  \\
\bottomrule
\end{tabular}
}\normalsize
\caption{The supersymmetric cones of flows for all values of $\omega$. The table shows the number $\#$ of cones dominated by an IR point with supersymmetry and bosonic symmetry $\cN$, $G$. The boundaries of the cones are also shown: 'G$_2$' stands for limiting flows with UV or IR endpoints in a G$_2$ point, and 'direct' stands for the direct flow from SO(8) in the UV to the corresponding point $G$ in the IR.}
\label{Table:cones}  
\end{table*}


\section{Supersymmetric RG flows between fixed points}\label{sec.flows}

In this section, we will discuss supersymmetric  domain wall solutions interpolating between the AdS fixed points of the $\omega$-deformed, SO(8)-gauged, $\cN=8$ supergravities \cite{Dall'Agata:2012bb} with at least SU(3) invariance that were classified in \cite{Borghese:2012zs}. As we will discuss, these solutions should holographically describe RG evolution between different conformal phases of the dual CFTs caused by UV-relevant deformations. Before dealing with the flows themselves, we first recall some relevant features of the SU(3)-invariant sector of the supergravity.

\subsection{The SU(3)-invariant sector} \label{sec:SU3inv}

The theory that results from truncating all, bosonic and fermionic, fields charged under SU(3) $\subset$ SO(8) and SU(3) $\subset$ SU(8), respectively, out of the full $\omega$-deformed $\cN=8$ supergravity \cite{Dall'Agata:2012bb} and retaining just SU(3) singlets was constructed in \cite{Borghese:2012zs}. This smaller sector is described by $\cN=2$ supergravity coupled to one vector and one hypermultiplet, with an abelian, compact, U(1)$^2$ gauging in the hypermultiplet sector. The corresponding six real scalars are coordinates on a
\begin{eqnarray} \label{ScalMan}
\frac{\textrm{SU}(1,1)}{\textrm{U}(1)} \times  \frac{\textrm{SU}(2,1)}{\textrm{SU}(2) \times \textrm{U}(1)} \,   
\end{eqnarray}
subspace of E$_{7(7)}/$SU(8) and can be conveniently parametrised in terms of a complex coordinate $z$ on the first, special-K\"ahler factor of (\ref{ScalMan}), and two complex coordinates $(\zeta_1 , \zeta_2) \equiv q^u$, $u=1,\ldots, 4$, on the second, quaternionic-K\"ahler factor.

The bosonic sector of this model is described by the lagrangian
\begin{eqnarray} \label{electricN=2action}
{\cal L}& = & \tfrac{1}{2} R *1 + g_{z\bar z} dz \wedge * d\bar z +h_{uv} Dq^u \wedge * Dq^v 
-V*1 \nonumber \\
&& +\tfrac{1}{2} \textrm{Im} \left(\cN_{IJ} \right)  F^{ I} \wedge *F^{J }
+\tfrac{1}{2} \textrm{Re} \left(\cN_{IJ} \right) F^{I} \wedge F^{J} \, ,
\end{eqnarray}
where the ($\omega$-independent) metrics of the non-linear-sigma-model-type kinetic terms for the scalars can be chosen to be
\begin{eqnarray} \label{ds2VM}
ds^2 = g_{z \bar z} dz d\bar z \equiv   \frac{3 dz d\bar z }{(1- |z|^2)^2} \, 
\end{eqnarray}
and
\begin{eqnarray} \label{ds2HM}
ds^2 = h_{uv} dq^ u dq^ v \equiv \frac{d \zeta_1 d\bar{\zeta_1} +  d \zeta_2 d\bar{\zeta_2}}{1-|\zeta_1|^2-|\zeta_2|^2} +  \frac{\big( \zeta_1 d \bar{\zeta_1} + \zeta_2 d \bar{\zeta_2} \big) \big( \bar{\zeta_1} d \zeta_1 + \bar{\zeta_2} d \zeta_2 \big) }{\big(1-|\zeta_1|^2-|\zeta_2|^2\big)^2} \, ,
\end{eqnarray}
on each factor of (\ref{ScalMan}). Out of the six real scalars of the model, only the four neutral under the gauge group enter the scalar potential $V$. These are the vector multiplet scalar $z$ and a combination of the hyperscalars that can be taken to be 
\begin{equation}
\label{zeta12}
\zeta_{12}\,\equiv\,\frac{|\zeta_{1}|+i|\zeta_{2}|}{1+\sqrt{1-|\zeta_{1}|^{2}-|\zeta_{2}|^{2}}} \ ,
\end{equation}
so that $(z, \zeta_{12})$ parametrise two copies of the Poincar\'e disk. The  potential can then be written as
\begin{eqnarray} \label{PotFromSuperPot}
V =  \,\,  2 g^2 \bigg[ \frac{4}{3} \, (1 - |z|^{2})^{2} \,  \frac{\partial |\cW |}{\partial z} \frac{\partial | \cW |}{\partial \bar z}  + (1 -  |\zeta_{12}|^{2} )^{2} \, \frac{\partial | \cW |}{\partial \zeta_{12}} \frac{\partial | \cW |}{\partial \bar \zeta_{12}}  - 3 |\cW|^{2}  \bigg] \, ,
 \end{eqnarray}
where $g$ is the SO(8) gauge coupling constant, and the complex, non-holomorphic, $\omega$-dependent quantity ${\cal W}$ is such that $|{\cal W}|$ plays the role of a superpotential for $V$. It is explicitly given by
\begin{eqnarray} 
\label{SU(3) omega superpotential+}
\cW =  (1 - |z|^{2})^{-3/2} \, (1 - |\zeta_{12}|^{2} )^{-2} \left[ (e^{2i \omega} + z^{3}) \, (1 + \zeta_{12}^{4}) + 6 \, z \, (1 + e^{2i \omega} z) \, \zeta_{12}^{2} \right] \, . 
\end{eqnarray}

We will not need to fully specify either the ($\omega$-dependent) gauge kinetic metric ${\cal N}_{IJ}$ or the hyperscalar covariant derivatives $Dq^u$, as these will not be needed for our analysis, and refer instead to \cite{Borghese:2012zs} for the details. We will only mention here that, in a dyonic frame with respect to the $\omega = 0$ electric/magnetic duality frame, these covariant derivatives do involve the electric, $A^I$, and magnetic $\tilde{A}_I$, $I=0,1$, graviphoton and vector in the vector multiplet as 
\begin{eqnarray} \label{covDers2}
D q^u = d q^u -  \Big( (A^0 \cos\omega -\tilde{A}_0 \sin\omega ) k_1^u + (A^1 \cos\omega -\tilde{A}_1 \sin\omega ) k_2^u \Big)  \, ,
\end{eqnarray}
with $k_1$, $k_2$ the U(1)$^2$ Killing vectors of SU(2,1)/(SU(2)$\times$U(1)). 

In the $\omega =0$ limit, the lagrangian (\ref{electricN=2action}) correctly reproduces that \cite{Bobev:2010ib} of the SU(3)-singlet sector of the original SO(8)-gauged supergravity \cite{deWit1}, and the potential (\ref{PotFromSuperPot}) accordingly recovers the critical structure studied by Warner in \cite{Warner:1983vz}. All vacua are AdS, and the supersymmetric ones, which are critical points of $|\cW|$ with $\cW$ in (\ref{SU(3) omega superpotential+}) at $\omega=0$, have supersymmetry and bosonic symmetry ($\cN = 8$, SO(8)),  ($\cN = 1$, G$_2$) and ($\cN = 2$, SU(3)$\times$U(1)). In the scalar parametrisation $(z, \zeta_{12})$ that we are using, these $\omega = 0$ critical points arise with multiplicities 1, 2 and 1 in the $z$ disk and and 1, 4 and 2 in the $\zeta_{12}$ disk. The fact that the multiplicities in the latter disk double those in the first for all critical points in the SU(3) sector, except  ($\cN = 8$, SO(8)), for any zero or non-zero value of $\omega$, is a generic feature of this parametrisation: a critical point other than the maximally supersymmetric point at $z=\zeta_{12}=0$ is specified by a position $z$ and two, equivalent, values of $\zeta_{12}$. We can thus refer to the {\it geometric multiplicity} of a critical point to be its multiplicity in the $z$ disk. On the other hand, critical points with geometric multiplicities greater than one may still need to be identified on physical grounds. Each critical point has two distinguishing features that carry physical information about the dual CFT$_3$: the cosmological constant  and the mass spectrum, which are related holographically to the free energy and the spectrum of conformal dimensions, respectively. At $\omega =0$, critical points with different supersymmetry and bosonic symmetry all have different spectra and different cosmological constants, and thus characterise distinct superconformal phases of the dual field theory. The two ($\cN = 1$, G$_2$) points, however, have both the same spectrum and the same cosmological constant, and must be identified under a $\mathbb{Z}_2$ transformation of the scalar manifold (\ref{ScalMan}) to be physically equivalent. These G$_2$ points thus actually have {\it physical multiplicity} one. This is the right interpretation from the dual field theory \cite{Bobev:2009ms}.

At finite $\omega$, the vacuum structure of this model was investigated in \cite{Borghese:2012zs}, although the identifications below were not considered there. The supersymmetric vacua are now critical points of $|\cW|$ with $\cW$ in (\ref{SU(3) omega superpotential+}) for generic $\omega$. The critical points above remain critical points at $\omega \neq 0$, although their location in the scalar manifold (\ref{ScalMan}) and the value of the cosmological constant at each point changes with $\omega$ (except for the SO(8) point, which remains with a fixed cosmological constant at the origin of moduli space). In particular, the two G$_2$ points develop different $\omega$-dependent cosmological constants for $0< \omega \leq \frac{\pi}{8}$, so they must now be considered physically different critical points in this range of $\omega$. Further, three more AdS critical points arise at $\omega \neq 0$, with  ($\cN = 1$, G$_2$), ($\cN = 2$, SU(3)$\times$U(1)) and ($\cN = 1$, SU(3)) symmetry, all of them with $\omega$-dependent positions in scalar space. The first two display the same, $\omega$-independent, mass spectra than the $\omega=0$ points with the same symmetries,  but have different, $\omega$-dependent, cosmological constants for $0<\omega < \frac{\pi}{8}$ and thus must be considered as different points. The SU(3) point is completely new and does not have an $\omega =0$ counterpart. The rightmost end, $\omega =\frac{\pi}{8}$, of the physically allowed range for $\omega$ (see footnote \ref{fnt:rangeomega}) is again special. Two of the three G$_2$ points and the two SU(3)$\times$U(1) points acquire the same cosmological constant so we respectively identify them by a $\mathbb{Z}_2$ transformation following the $\omega = 0$ intuition. As we will see later, these identifications will have consequences for our interpretation of the superconformal phases of the dual CFTs.

To summarise (see table  \ref{Table:SU(3)}), at $\omega = 0$ there exist three physically independent critical points with symmetries ($\cN = 8$, SO(8)),  ($\cN = 1$, G$_2$) and ($\cN = 2$, SU(3)$\times$U(1)). For $0< \omega < \frac{\pi}{8}$, there are seven physically independent critical points: one with symmetry ($\cN = 8$, SO(8)), three with  ($\cN = 1$, G$_2$), two with  ($\cN = 2$, SU(3)$\times$U(1)) and one with  ($\cN = 1$, SU(3)). Finally, when $\omega = \frac{\pi}{8}$, there are five physically independent critical points: one with symmetry  ($\cN = 8$, SO(8)), two with ($\cN = 1$, G$_2$), one with ($\cN = 2$, SU(3)$\times$U(1)) and one with ($\cN = 1$, SU(3)).

\begin{table*}\centering
\ra{1}
{\small
\begin{tabular}{@{}rcrccrccr@{}}\toprule
& \multicolumn{2}{c}{$\omega = 0$} & \phantom{abc}& \multicolumn{2}{c}{$\omega = \frac{\pi}{16}$} &
\phantom{abc} & \multicolumn{2}{c}{$\omega = \frac{\pi}{8}$}\\ \cmidrule{2-3} \cmidrule{5-6} \cmidrule{8-9}
& g.m. & $V_*/g^2$ && g.m. & $V_*/g^2$ &&  g.m. & $V_*/g^2$\\ \midrule
$\cN=8$, SO(8) 
       & 1  & $-6$ && 1  & $-6$ && 1  & $-6$\\[14pt]
$\cN=2$, SU(3)$\times$U(1)  
       & 1 & $-7.794$ &&  1  &  $-7.912$  && 2  & $-8.354$ \\
       &      &                &&   1  & $-9.672$ &&        &                 \\[14pt]
$\cN=1$, G$_2$
       & 2  & $-7.192$ &&  1  &  $-7.075$  && 1  & $ -7.040$ \\
       &       &                &&   1  & $-7.436 $ &&  2    &   $ -7.943 $              \\
       &     &                &&   1  & $  -9.264 $ &&        &                 \\[14pt]
$\cN=1$, SU(3)
       &     &                 &&  1  &  $-11.353$  && 1  & $-10.392$ \\
\bottomrule
\end{tabular}
}\normalsize
\caption{The critical points in the SU(3) sector at both ends of the physically allowed range of $\omega$ and at an intermediate value. For each point,  the geometric multiplicity, g.m., and the $\omega$-dependent value $V_*$ of the cosmological constant are shown. Each entry has physical multiplicity 1, that is, corresponds to a physically distinct phase. All points with the same symmetry $\cN$, G, have the same $\omega$-independent mass spectrum (see table \ref{Table:Masses}).}
\label{Table:SU(3)}  
\end{table*}

\subsection{Supersymmetric flows} \label{subsec:susyflows}

We want to study  supersymmetric domain walls interpolating between the supersymmetric critical points within the SU(3)-invariant sector of  each particular supergravity at fixed $\omega$, that preserve at least $\cN=1$ supersymmetry along the flow. We thus take a  domain wall ansatz for the four-dimensional spacetime metric 
\begin{equation} \label{eq:DW}
ds^2 = e^{2A(r)} \left( -dt^2+dx^2+dy^2 \right) + dr^2 \ ,
\end{equation}
and assume that the only coordinate dependence of the scalars $(z,\zeta_{12})$ is on the radial coordinate $r$. Of course, all functions $(A, z, \zeta_{12})$ will also depend parametrically on $\omega$. The BPS flow equations then follow from the vanishing of the fermionic supersymmetry variations that leave the action (\ref{electricN=2action}) invariant and, in this context, reduce to
 \begin{equation} \label{eq.domainwall}
 \partial_r A = -\sqrt{2}\,g |\cW| \ ,
 \end{equation}
 for the metric function and
\begin{equation} \label{eq.BPSs}
\partial_r z = \frac{2\sqrt{2}\,g}{3} \,(1-|z|^2)^2\frac{\partial |\cW|}{\partial \bar z} \ , \qquad \partial_r \zeta_{12} = \frac{g}{\sqrt{2}}\,(1-|\zeta_{12}|^2)^2 \frac{\partial |\cW|}{\partial \bar\zeta_{12}} \ , 
\end{equation}
for the complex scalars. The superpotential $|\cW|$ is the absolute value of (\ref{SU(3) omega superpotential+}).

At a fixed point $z=z_*$ and $\zeta_{12}=\zeta_*$, we have $\partial_r z=\partial_r\zeta_{12}=0$ and
\begin{equation}
\partial_r A = -\sqrt{2}\, g |\cW_*| \Rightarrow A=  \frac{r}{L_*} \ ,
\end{equation}
where $L_*=-( \sqrt{2}\, g |\cW_*|)^{-1}$ is the radius of the AdS$_4$ spacetime associated to the fixed point (this can be seen by e.g. changing coordinates to $\eta=L_* \,e^{-r/L_*}$ in the generic domain wall metric (\ref{eq:DW})). The (Poincar\'e patch of) AdS space itself can indeed be regarded as a domain wall with metric function $A$ linear in $r$. The UV (IR) region arises as $r\to\infty$ ($r\to -\infty$) in the domain wall (\ref{eq:DW}), or at $\eta=0$ ($\eta\to\infty$) in the  Poincar\'e radius. In the parametrisation for the scalar fields that we are using, the SO(8), $\cN=8$ point is locked at the origin $z= \zeta_{12}=0$, independently of $\omega$. The value $|\cW_*|$ of the superpotential at this central point is also $\omega$-independent and gives rise to the highest (i.e., lowest in absolute value), cosmological constant among the supersymmetric critical points with at least SU(3) symmetry (see table \ref{Table:SU(3)}). This is consistent with the expectation that the SO(8) point  always serves as the UV origin of  flows driven by mass deformations of the dual CFT that we discuss below. At the SO(8) point, $|\cW_{*}|=1$ and, accordingly, $L_{\textrm{SO(8)}}=-(\sqrt{2}\, g)^{-1}$. We henceforth set $g=-1/\sqrt{2}$ and thus $L_{\textrm{SO(8)}}=1$, namely, we measure the AdS radii of all other fixed points with respect to the AdS radius of the $\cN=8$ point.

To obtain the domain wall solutions between different critical points at a fixed $\omega$, we integrate numerically the BPS equations \eqref{eq.domainwall}, \eqref{eq.BPSs} with suitable boundary conditions specified by the physics. More concretely, we first obtain linearised solutions to \eqref{eq.BPSs} around each critical point $(z_*, \zeta_*)$. These turn out to be of the form
\begin{equation}\label{eq.perturb}
z  = z_* + \sum_{i=1}^4 z_{0 i } \, e^{-\tilde{\Delta}_i r /L_*} \ , \quad \zeta_{12}  = \zeta_{*}  +  \sum_{i=1}^4 \zeta_{0i}  \, e^{-\tilde{\Delta}_i  r/L_*}  \ . 
\end{equation}
Here, $z_{0i}$ and $\zeta_{0i}$ are linear combinations of the four, real, independent integration constants arising from each of the four real differential equations contained in  \eqref{eq.BPSs}, and each $\tilde{\Delta}_i$, $i=1,\ldots, 4$, 
is a root, either the long, $\Delta_+$, or the short one, $\Delta_-$, of the quadratic equation  
\begin{eqnarray} \label{eq:MassDimRel}
(m L_*)^2 = \Delta (\Delta  -3) \, , 
\end{eqnarray}
for each eigenvalue $m_i^2$ of the scalar mass matrix at the critical point $(z_* , \zeta_{*})$. In particular, we find that the $\tilde{\Delta}$'s are independent of $\omega$, in agreement with the $\omega$-independence of the mass spectrum of the supersymmetric points with at least SU(3) symmetry \cite{Borghese:2012zs}, and find them to be compatible with the masses quoted in that reference. In table \ref{Table:Masses} we have summarised the masses and dimensions associated to the scalars $z$, $\zeta_{12}$, and have highlighted the values of the $\tilde \Delta$'s that appear in (\ref{eq.perturb}) in bold. The spectra can be arranged in OSp$(4|\cN)$ multiplets as corresponds to $\cN$-supersymmetric points. From the data of the table, it is straightforward to organise the $\cN=1$ spectra into OSp$(4|1)$ chiral multiplets, but the arrangement into OSp$(4|2)$ multiplets for SU(3)$\times$U(1) involves all fields of the $\cN=2$ lagrangian (\ref{electricN=2action}). See \cite{Borghese:2012zs} for the details.

\begin{table*}\centering
\ra{1.1}
{\tiny
\begin{tabular}{@{}rrrcrrcrrcrr@{}}
\toprule
& \multicolumn{2}{c}{Eigenvalue 1} & \phantom{abc}& \multicolumn{2}{c}{Eigenvalue 2} &
\phantom{abc} & \multicolumn{2}{c}{Eigenvalue 3} &
\phantom{abc} & \multicolumn{2}{c}{Eigenvalue 4} \\ \cmidrule{2-3} \cmidrule{5-6} \cmidrule{8-9} \cmidrule{11-12}
& an. & num.   && an. & num.  &&  an. & num.  &&  an. & num.  \\ \midrule
$\cN=8$, SO(8) \\
   $(mL_*)^2$    & $-2$ & $-2.$  && $-2$ & $-2.$  && $-2$ & $-2.$ && $-2$ & $-2.$ \\
  $\Delta_+$    & 2 & 2.  && 2 & 2.  && 2 & 2. && 2 & 2. \\
  $\Delta_-$    & $\mathbf{1}$ & $\mathbf{1.}$  && $\mathbf{1}$ & $\mathbf{1.}$ && $\mathbf{1}$ & $\mathbf{1.}$ && $\mathbf{1}$ & $\mathbf{1.}$ \\[10pt]
$\cN=2$, SU(3)$\times$U(1)  \\
   $(mL_*)^2$    & $3-\sqrt{17}$ & $-1.123$  && $2$ & $2.$  && $2$ & $2.$ && $3+\sqrt{17}$  & $7.123$ \\
  $\Delta_+$    & $\mathbf{\frac{1+\sqrt{17}}{2}}$  & $\mathbf{2.562}$  && $\mathbf{\frac{3+\sqrt{17}}{2}}$  & $\mathbf{3.562}$  && $\frac{3+\sqrt{17}}{2}$  & 3.562&& $\frac{5+\sqrt{17}}{2}$ & 4.562 \\
  $\Delta_-$    & $\frac{5-\sqrt{17}}{2}$ & 0.438  && $\frac{3-\sqrt{17}}{2}$  & $-0.562$ && $\mathbf{\frac{3-\sqrt{17}}{2}}$  & $\mathbf{-0.562}$ && $\mathbf{\frac{1-\sqrt{17}}{2}}$ & $\mathbf{-1.562}$ \\[10pt]
  $\cN=1$, G$_2$  \\
   $(mL_*)^2$    & $-\frac{11+\sqrt{6}}{6}$ & $-2.242$  && $\frac{-11+\sqrt{6}}{6}$ & $-1.425$  && $4-\sqrt{6}$ & $1.551$ && $4+\sqrt{6}$  & 6.450 \\
  $\Delta_+$    & $2-\frac{1}{\sqrt{6}}$ & 1.592  && $2+\frac{1}{\sqrt{6}}$  & 2.408  && $\mathbf{1+\sqrt{6}}$  & $\mathbf{3.450}$ && $2+\sqrt{6}$  & 4.450 \\
  $\Delta_-$    & $\mathbf{1+\frac{1}{\sqrt{6}}}$ & $\mathbf{1.408}$  && $\mathbf{1-\frac{1}{\sqrt{6}}}$  & $\mathbf{0.592}$ && $2-\sqrt{6}$   & $-0.450$ && $\mathbf{1-\sqrt{6}}$   & $\mathbf{-1.450}$ \\[10pt]
  $\cN=1$, SU(3)  \\
   $(mL_*)^2$    & $4-\sqrt{6}$ & $1.551$  && $4-\sqrt{6}$ & $1.551$  && $4+\sqrt{6}$  & 6.450 && $4+\sqrt{6}$  & 6.450 \\
  $\Delta_+$    &$\mathbf{1+\sqrt{6}}$  & $\mathbf{3.450}$  && $\mathbf{1+\sqrt{6}}$  & $\mathbf{3.450}$  && $2+\sqrt{6}$  & 4.450 && $2+\sqrt{6}$  & 4.450 \\
  $\Delta_-$    & $2-\sqrt{6}$   & $-0.450$ && $2-\sqrt{6}$   & $-0.450$ &&  $\mathbf{1-\sqrt{6}}$   & $\mathbf{-1.450}$ && $\mathbf{1-\sqrt{6}}$   & $\mathbf{-1.450}$ \\
 \bottomrule
\end{tabular}
}\normalsize
\caption{The ($\omega$-independent) scalar mass spectrum about all critical points of the SU(3) invariant sector. For each eigenvalue $(mL_*)^2$ of the mass matrix, the long, $\Delta_+$, and short, $\Delta_-$, roots of equation (\ref{eq:MassDimRel}) are given, both analytically and numerically. Entries in bold correspond to the values $\tilde{\Delta}$ that appear in the linearised flow solutions (\ref{eq.perturb}).}
\label{Table:Masses}  
\end{table*}

Having obtained the linearised solution about each critical point, we then integrate the full solution. According to the fall-off (\ref{eq.perturb}), only modes with $\tilde \Delta >0$ or $\tilde \Delta <0$ give rise to regular solutions at a UV ($r\to\infty$) or IR ($r\to-\infty$) fixed point. Boundary conditions can be chosen in order to select the appropriate modes. Specifically, we find the radial integration to be under much better numerical control if we start to integrate from the IR towards the UV. Accordingly, we specify boundary conditions so that $z_{0i} = \zeta_{i0} = 0$ for all modes $i=1, \ldots, 4$ with $\tilde{\Delta}_i > 0$ at an IR fixed point, in order to have a smooth incoming flow with only $\tilde{\Delta} <0$ modes in the IR. Then, we numerically shoot in order to find the radial profile of the scalars $(z, \zeta_{12})$ and the metric function $A$, and determine the UV end of the flow.

The constants $\tilde{\Delta}$ not only govern the behaviour of a domain wall near a critical point, but also carry information on the type of RG flow that the wall is interpreted to describe holographically. Firstly, recall that deformations of a UV CFT$_3$ by relevant  operators (those with scaling dimension $\Delta < 3$) will trigger an RG flow which, if it ends on another conformal phase, will be driven into this IR phase by an irrelevant ($\Delta > 3$) deformation. Secondly, recall the holographic prescription that a scalar in the $D=4$ bulk with mass $m^2$ is dual to a field theory operator of dimension $\Delta$ given by the longest root, $\Delta=\Delta_+$, of the quadratic equation (\ref{eq:MassDimRel}). Finally, recall that, for masses outside of the range $-\tfrac{9}{4}<m^2 L_*^2\leq -\tfrac{5}{4}$, a (non-normalisable) fall-off $\tilde \Delta = \Delta_-$ and a (normalisable) fall-off $\tilde \Delta = \Delta_+$ in the linearised flow solutions  (\ref{eq.perturb}) respectively correspond to deformations or vevs in the dual field theory. 

For the SO(8) point, all four $\tilde \Delta$'s are equal to +1, leading to linearised flow equations
\begin{equation}\label{eq.UVperturb2}
z  = z_0 \, e^{- r} \ , \qquad \zeta_{12}  = \zeta_0 \, e^{- r}  \ , 
\end{equation}
and implying that only domain walls with this point at its UV origin can be regular. The subtlety in this case is that all masses at this point lie within the range above, so both possible fall-offs correspond to alternative quantisations. For all other critical points in the SU(3) sector, the positive (in UV points) and negative (in IR points)  $\tilde \Delta$'s always turn out to correspond to $\Delta_-$ roots, and thus to non-normalisable fall-offs responsible for deformations of the dual field theory lagrangian. Rather remarkably, these deformations further turn out to be always relevant (in UV points) and irrelevant (in IR points). This thus leads to a perfectly consistent holographic description of a web of RG flows of the dual large $N$ field theories among the phases with at least SU(3) invariance triggered by suitable deformations. Observe from table \ref{Table:Masses} that all critical points other than SO(8) have negative $\tilde \Delta$'s in their spectrum and can therefore serve as IR points. Finally, all flows of course proceed from points with higher (in the UV) to lower (in the IR) cosmological constant.

\subsection{Flows at $\omega = 0$} \label{w=0Flows}

The holographic RG flows between critical points with at least SU(3) invariance of the 
SO(8)-gauged supergravity of \cite{deWit1} were exhaustively constructed and interpreted in \cite{Bobev:2009ms}. Here we review their results, both as a check on our numerics and to set the stage for the $\omega \neq 0$ situation. This will turn out to be a more elaborate version of the $\omega = 0$ case. 

The $\omega = 0$  fixed points are located at
\begin{align}
\label{eq:SO8pointw=0}
\textrm{SO(8)} \rightarrow & \quad z = \zeta_{12} = 0 \ , \\
\label{eq:G2pointw=0}
\textrm{G}_2 \rightarrow & \quad   z= \pm \zeta_{12}  = \frac{3+\sqrt{3}-3^{1/4}\sqrt{10}}{4} \left(1\pm i\,3^{-1/4}\sqrt{2+\sqrt{3}}\right) \ , \\
\label{eq:SU3U1pointw=0}
\textrm{SU(3)}\times \textrm{U(1)} \rightarrow &  \quad  z = 2 - \sqrt{3} \ , \qquad \zeta_{12} = \pm i\, \big( \sqrt{3} -\sqrt{2} \big) \ , 
\end{align}
with AdS$_4$ radii
\begin{eqnarray} \label{eq:AdSradiiw=0}
L_{\textrm{SO(8)}} = 1\, , \quad L_{\textrm{G}_2}=\frac{1}{3^{1/8}} \frac{5}{6} \left( \frac{5}{2} \right)^{1/4} \, , \quad  L_{\textrm{SU(3)}\times \textrm{U(1)}} = \frac{2}{3^{3/4}}  \, ,
\end{eqnarray}
leading to the numerical values for the cosmological constants quoted in table \ref{Table:SU(3)}. The linearised solutions of the BPS equations (\ref{eq.BPSs}) around these points are (\ref{eq.perturb}) with $\tilde \Delta$'s given by the entries in bold in table \ref{Table:Masses} and the corresponding $L_*$ in (\ref{eq:AdSradiiw=0}). As we mentioned at the end of section 
\ref{subsec:susyflows}, only the G$_2$ and SU(3)$\times$U(1) points can arise as IR points.

\begin{figure}[tb]
\begin{center}
\includegraphics[scale=0.45]{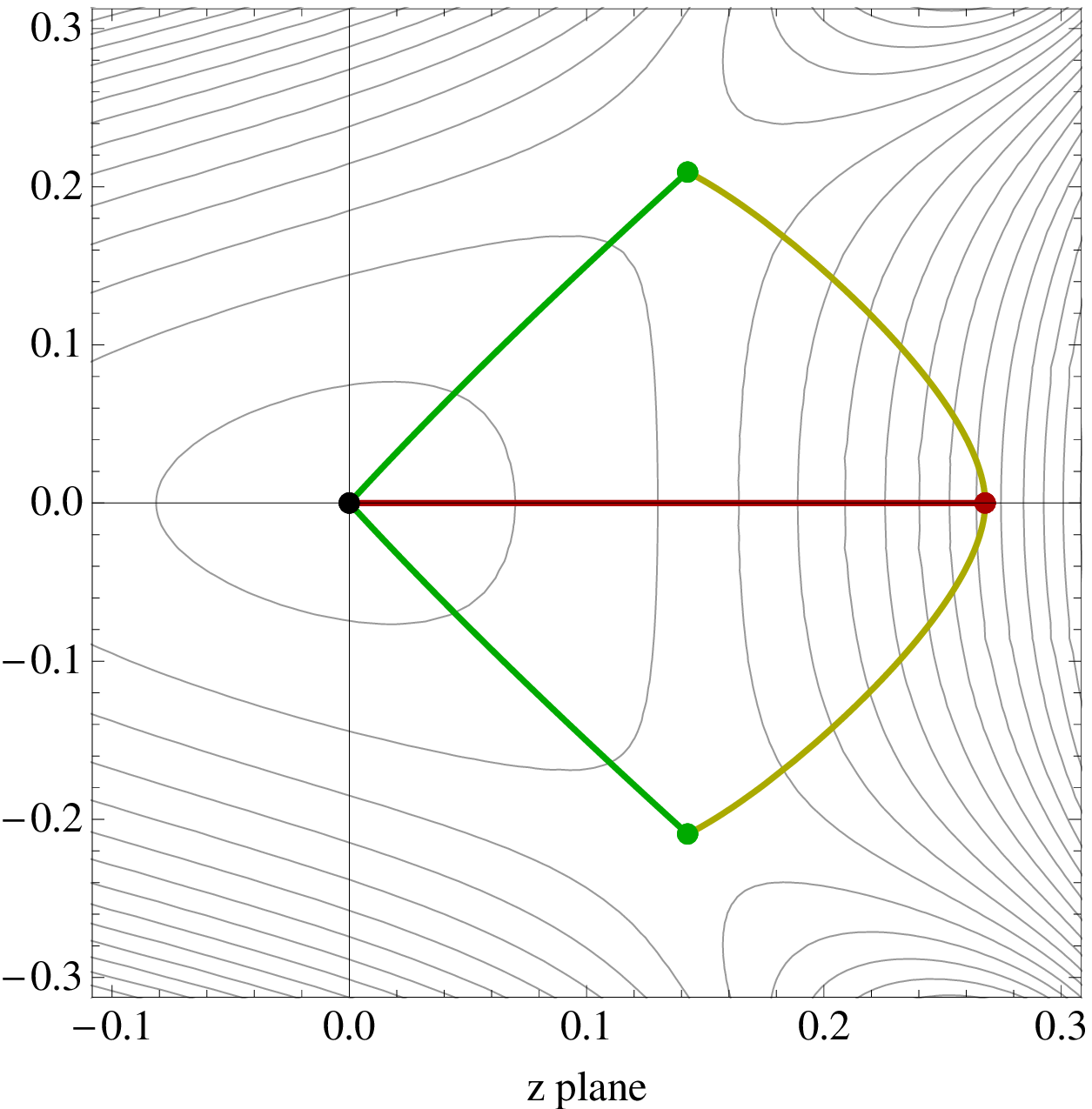}
\includegraphics[scale=0.45]{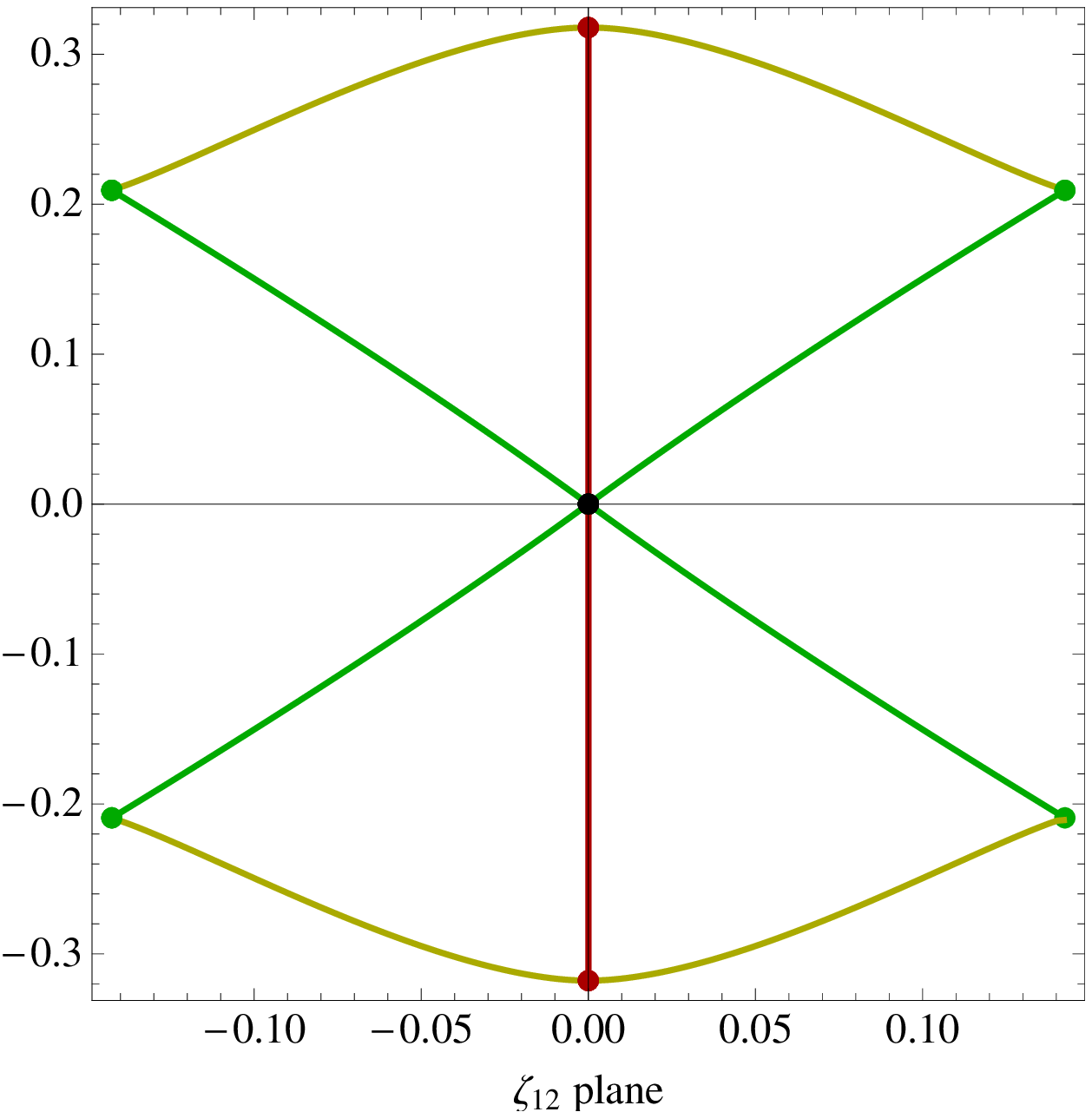}
\caption{The numerically-generated trajectories of the $\omega = 0 $ unique and direct flows in the $z$ (left) and $\zeta_{12}$ (right) disks. The SO(8), G$_2$ and SU(3)$\times$U(1) points are shown in black, green and red. Both G$_2$ points are equivalent and so are the flows with them as IR (green) or UV (yellow) endpoint.}\label{fig.wis0}
\end{center}
\end{figure}

Consider first the G$_2$ points with geometric multiplicity 2 (corresponding to the $\pm$ signs inside the bracket of the right hand side of (\ref{eq:G2pointw=0})) as IR points. From table \ref{Table:Masses}, only one mode has a negative $\tilde \Delta$, given by $\tilde{\Delta} = -1.450$. In order to choose boundary conditions so that only this mode is active, we set to zero three of the four real integration constants upon which the $z_{0i}, \zeta_{0i}$ in (\ref{eq.perturb}) depend. The remaining real constant can be fixed by a radial coordinate redefinition, which thus leaves (with the specified boundary conditions) a unique, regular flow with G$_2$ as the IR endpoint. Numerically shooting, we verify that this domain wall lands at the SO(8) point, which therefore dominates the UV physics. From table \ref{Table:Masses}, the holographic interpretation of this domain wall is clear \cite{Bobev:2009ms}. The perturbation $\tilde{\Delta} = -1.450$ corresponds to the non-normalisable mode, $\tilde{\Delta} = \Delta_-$, of an IR-irrelevant operator of conformal dimension $\Delta_+ = 4.450$. This SO(8) to G$_2$ domain wall thus describes holographically the RG flow between superconformal phases with those symmetries, triggered by a mass deformation of the UV field theory, preserving $\cN=1$ supersymmetry and G$_2$ bosonic symmetry along the flow, and landing into the IR driven by the above irrelevant deformation. The green curves in figure \ref{fig.wis0} correspond to the trajectory in the $z$, $\zeta_{12}$ disks of the numerically integrated flow (note that the two curves in the figure correspond to the same physical flow). In this and all other figures in this section, we have used contour lines in the $z$ disk (we have omitted them in the $\zeta_{12}$ disk) corresponding to a projection of the superpotential  $|\cW|$ to the $z=\zeta_{12}$ plane. 
This projection is consistent with a G$_2$-invariant truncation of the theory, and therefore the contours only reflect accurately  the G$_2$ fixed points, denoted with green dots. The SO(8), SU(3)$\times$U(1) and, in the next sections, the SU(3) points are represented with black, red and blue dots, respectively. 
Incidentally, note that for this flow and its $\omega \neq0$ counterparts of the next sections, $z = \zeta_{12}$ and, hence, G$_2$ invariance, holds all along the flow.

Let us next focus on the SU(3)$\times$U(1) point. As shown in table \ref{Table:Masses}, now two modes have negative $\tilde{\Delta}$'s. Choosing boundary conditions so that only these modes are active sets to zero two out of the four real integration constants. As above, one of the remaining constants can be fixed by a radial shift and, consequently, we are left with a one-parameter family (or {\it cone}, following the terminology of \cite{Bobev:2009ms}) of flows ending in the IR at the SU(3)$\times$U(1) point. We call this parameter $\lambda$. In agreement with \cite{Bobev:2009ms}, we find that there exists a preferred closed interval $[\lambda_-,\lambda_+]$ for the parameter.\footnote{For values of $\lambda$ outside this range, the domain wall solutions flow to the boundary, $|z|=1$, $|\zeta_{12}|=1$ of the Poincar\'e disks, that is, infinity of the corresponding upper half planes.} 

At either end of the interval, $\lambda = \lambda_-$ or $\lambda = \lambda_+$, the numerical integration towards to UV leads the domain wall toward the G$_2$ point in the fourth or first quadrant of the $z$ plane (corresponding to the $\pm$ signs in the expression for $z$ in (\ref{eq:G2pointw=0})), respectively, and their $\zeta_{12}$ companions. These walls thus interpolate between either G$_2$ point in the UV and the SU(3)$\times$U(1) point in the IR. In order to further interpret these flows, we again follow \cite{Bobev:2009ms}. Both $\lambda = \lambda_-$ and  $\lambda = \lambda_+$ flows reach the  SU(3)$\times$U(1) IR point with a combination of non-normalisable fall-offs $\tilde \Delta =  \Delta_- = -0.562$ and  $\tilde \Delta =  \Delta_- = -1.562$, thus corresponding holographically to insertions in the dual IR field theory lagrangian of irrelevant operators of dimensions $\Delta_+ = 3.562$ and $\Delta_+ = 4.562$. Remarkably and consistently enough, this interpretation also agrees from the UV point of view \cite{Bobev:2009ms}. If we now consider either G$_2$ point as the UV origin of a flow, only modes with positive $\tilde \Delta$ must be chosen in the linearised solution (\ref{eq.perturb}) of the BPS equations (\ref{eq.BPSs}). From table \ref{Table:Masses} we see that, of the three positive $\tilde \Delta$'s for the G$_2$ point, two ($\tilde \Delta = \Delta_- = 1.408$ and $\tilde \Delta = \Delta_- = 0.592$) correspond to non-normalisable fall-offs of modes of dimension $\Delta_+ = 1.592$ and $\Delta_+ = 2.408$, thus indeed corresponding to insertions of relevant operators into the dual UV G$_2$-symmetric field theory lagrangian that make the theory RG-flow into the SU(3)$\times$U(1) IR. These flows preserve SU(3) symmetry and $\cN=1$ supersymmetry, and correspond to the numerically generated yellow trajectories in figure \ref{fig.wis0}. Note that both G$_2$ points and both flows are physically equivalent.

For all values of $\lambda$ in the open interval $(\lambda_- , \lambda_+)$, we find that the UV endpoint of the family of domain walls corresponds to the $\cN=8$ SO(8) fixed point, in agreement with  \cite{Bobev:2009ms}. There exists a preferred $\lambda_0$ within the interval, $\lambda_- < \lambda_0  < \lambda_+$ corresponding to a {\it direct flow} between the SO(8) in the UV and the SU(3)$\times$U(1) point in the IR. For this domain wall, supersymmetry is enhanced to $\cN=2$ (within the full $\cN=8$ supergravity\footnote{\label{ftnt:N=1inN=2} As follows from the results of \cite{Hristov:2009uj,Louis:2012ux}, this flow can only be $\cN=1$ when considered as a solution of the $\cN=2$ subtruncation corresponding to the SU(3)-invariant sector.}) and the bosonic symmetry along the flow is the full SU(3)$\times$U(1) of the IR. This is the domain wall first constructed in \cite{Ahn:2000aq} and interpreted holographically in \cite{Benna:2008zy}. We have plotted the numerically generated trajectory of this direct flow in figure \ref{fig.wis0}: it corresponds to the red, straight line along the real axis in the $z$ disk (or the imaginary axis in the $\zeta_{12}$ disk). Figure \ref{fig.wis0} suggests that the red flow minimises the length of the trajectory between both endpoints, and we have numerically verified that this is indeed the case. Note that this statement is independent of the parametrisation $z$, $\zeta_{12}$ that we have used. In agreement with  \cite{Ahn:2000aq}, we find that only the mode with fall-off $\tilde \Delta = \Delta_- = -1.562$ drives the direct flow in the IR. Thus, the $\cN=2$, direct flow is (explicitly) being driven by only one mass deformation of the UV, SO(8) field theory lagrangian and lands into the SU(3)$\times$U(1) point driven by an irrelevant deformation of scaling dimension $\Delta_+ = 4.562$.

\begin{figure}[tb]
\begin{center}
\includegraphics[scale=0.45]{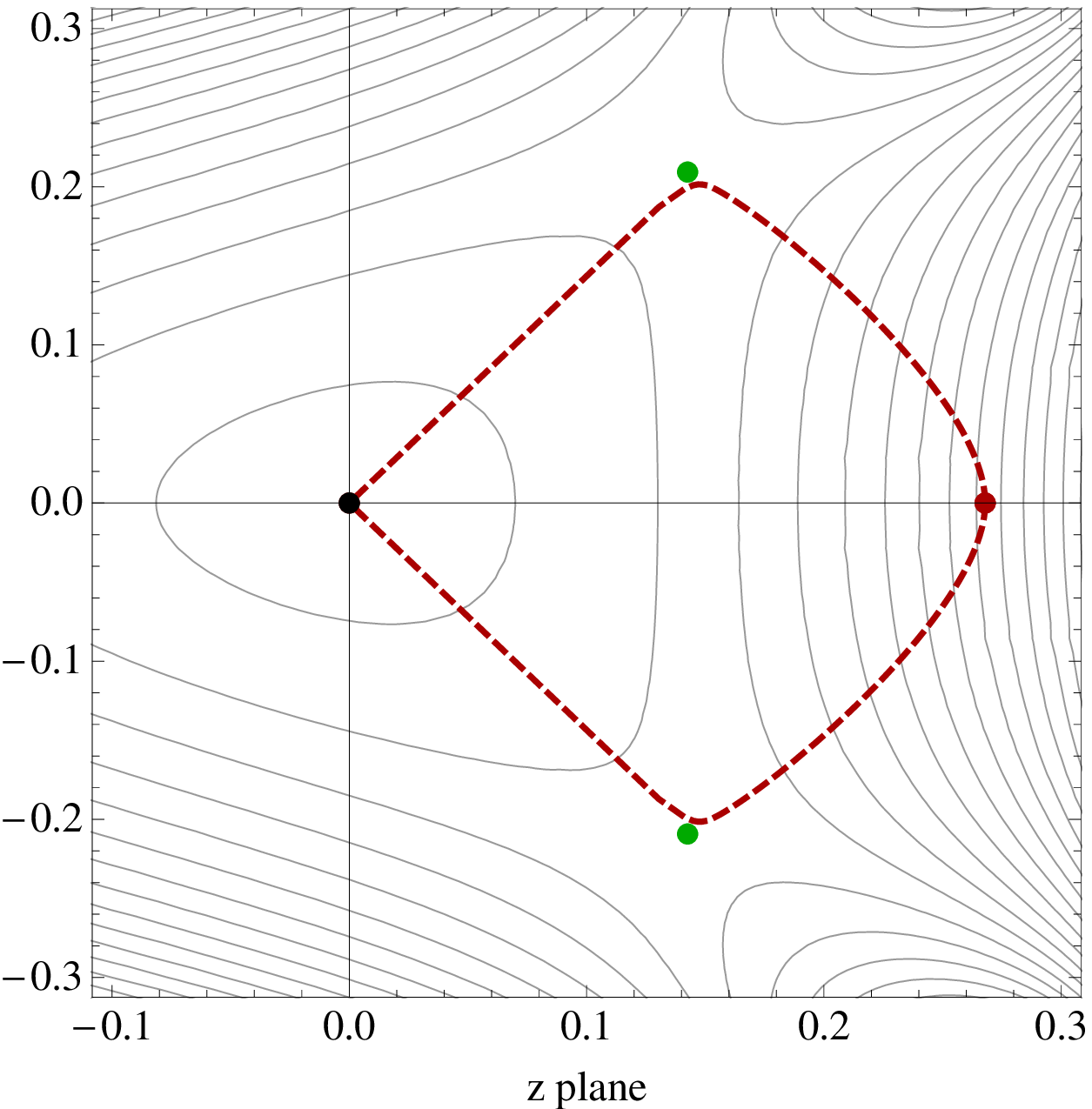}
\includegraphics[scale=0.45]{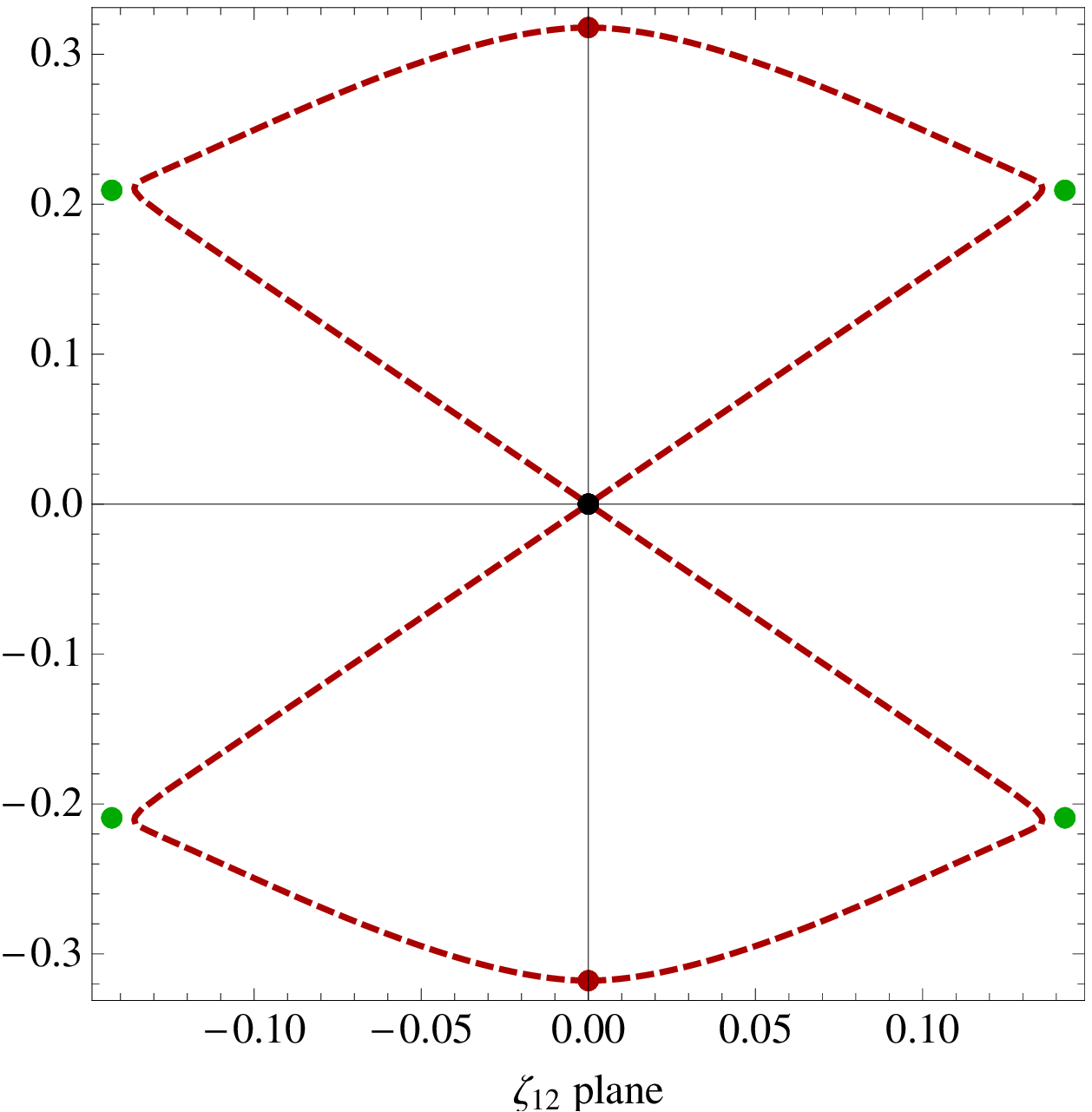}
\caption{Two (physically equivalent) flows in the $\omega=0$ SU(3)$\times$U(1) cone, following a path very close to the boundary of the cone dominated by the G$_2$ point.}\label{fig.wis0bis}
\end{center}
\end{figure}

More generally, for $\lambda$ in the open interval $(\lambda_- , \lambda_+)$ and different from $\lambda_0$, the corresponding flow is only $\cN=1$. It is driven by a relevant, mass deformation of the $\cN=8$, SO(8) UV phase into the $\cN=2$, SU(3)$\times$U(1) IR phase, where it lands driven by a combination of both irrelevant deformations of dimensions $\Delta_+ = 3.562$ and $\Delta_+ = 4.562$. This cone of flows is bounded, on the one hand, by the SO(8) in the UV to G$_2$ flow in the IR (the green curves in figure \ref{fig.wis0}), and by the G$_2$ in the UV to SU(3)$\times$U(1) flow in the IR (the yellow curves in the figure). In particular, there are flows that follow very closely these limiting flows, and pass arbitrarily near the G$_2$ fixed point without ever reaching it. For these solutions, the radial profiles of the scalars pass an arbitrarily large (radial) time close to the value corresponding to the G$_2$ fixed point solution. On the other hand, the cone is also bounded by the direct flow along the real axis. Indeed, observe that figures \ref{fig.wis0} and \ref{fig.wis0bis} are symmetric with respect to a $\mathbb{Z}_2$ reflection about the real axis. Both G$_2$ points are indeed physically equivalent, and so are flows mapped into each other by the  $\mathbb{Z}_2$ reflection. Although we have generated flow solutions that span the interior of the cone, as an example of these we only plot here, in figure \ref{fig.wis0bis}, two physically equivalent flows that proceed very closely to the boundary of the cone dominated by the G$_2$ point.

\subsection{Flows for $0< \omega < \frac{\pi}{8}$ }  \label{wIn0ToPion8}

As reviewed in section \ref{sec:SU3inv}, the superpotential $|\cW|$ has seven inequivalent extrema for generic, non-zero values of the electric/magnetic duality phase $\omega$. On the one hand, the four geometric extrema of the superpotential in the  parametrisation $(z, \zeta_{12})$ (of which only three are physically distinct when $\omega =0$), start changing their positions in the scalar manifold (except the SO(8) point). The cosmological constants for the two geometric G$_2$ extrema start to differ at non zero $\omega$, thus now rendering these two points physically inequivalent too. On the other hand, the extra three fixed points at finite $\omega$ can be thought of as being {\it imported} from the boundary $|z|=1$, $|\zeta_{12}|=1$ of the scalar manifold. Indeed, when $\omega$ is varied, the positions of these new three points in the $(z, \zeta_{12})$ disks follow a trajectory that tends to  $|z| \rightarrow 1$, $|\zeta_{12}| \rightarrow 1$ when $\omega \rightarrow 0$ \cite{Borghese:2012zs}. Although extrema with the same symmetries exhibit the same spectrum of masses, all seven points have different cosmological constants for $0 < \omega <\frac{\pi}{8}$, and thus must be regarded as physically inequivalent  in this interval for $\omega$.

\begin{figure}[tb]
\begin{center}
\includegraphics[scale=0.45]{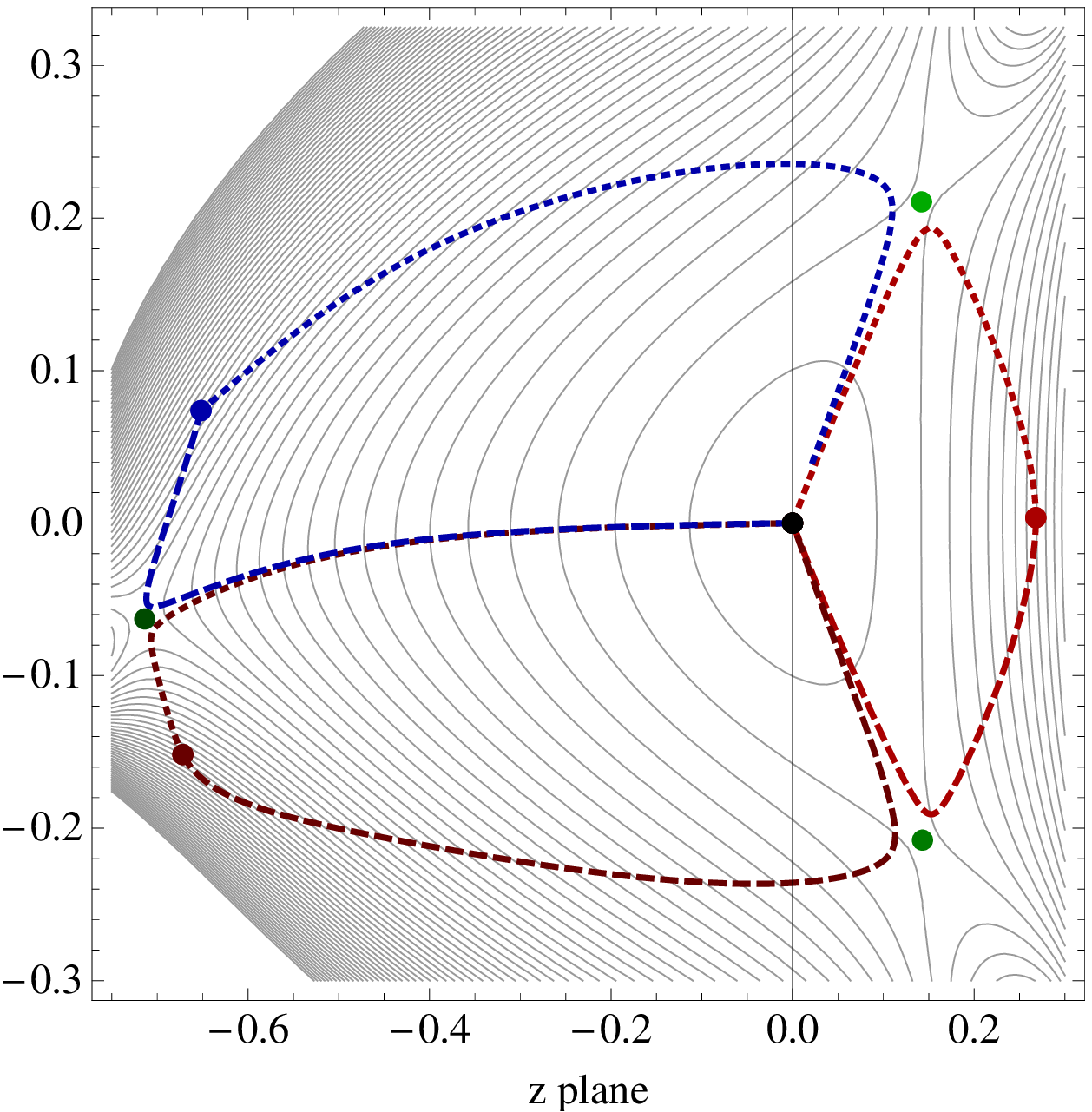}
\includegraphics[scale=0.45]{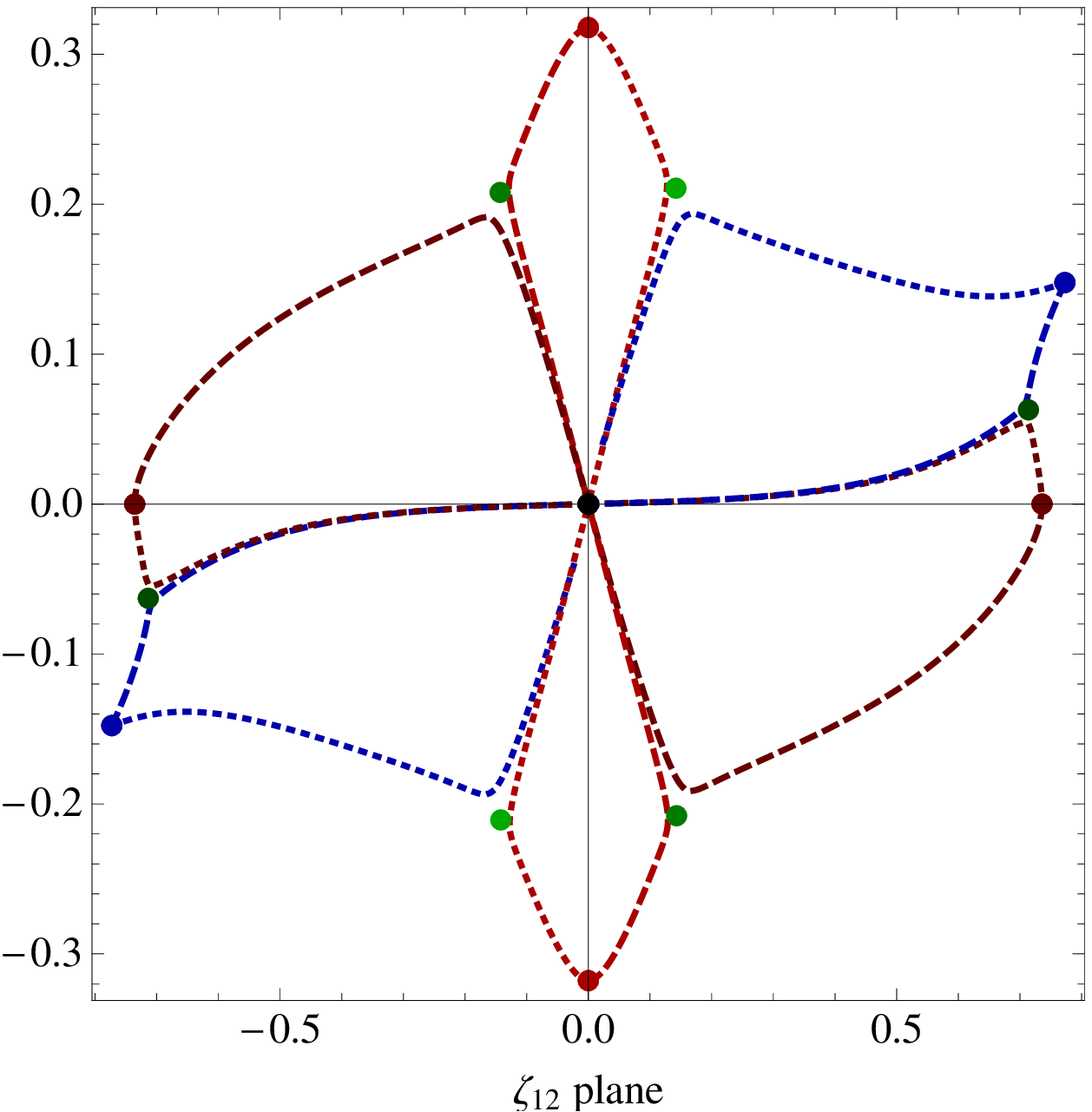}
\caption{Some flows at a small, non-zero value, $\omega=10^{-2}$, of the angle $\omega$. The blue dot corresponds to the new SU(3) point, and all other points are coloured as in previous plots. The plotted flows follow paths very close to the boundaries of the three physically inequivalent cones. Different colour shades and different dashings correspond to different physically distinct points and flows.}\label{fig.0p01bis}
\end{center}
\end{figure}

The dots in figure \ref{fig.0p01bis} correspond to the numerical location of the critical points for a small value of $\omega$. Referring to the $z$ plane in the left panel of the figure (the $\zeta_{12}$ plane in the right panel follows along), the position of the critical points in quadrants 1 and 4 is a small deviation from their $\omega = 0$ positions (compare with figure \ref{fig.wis0bis}). The new fixed points arise on the left hand side of the picture, in quadrants 2 and 3 of the $z$ plane. Points with SO(8), SU(3)$\times$U(1), G$_2$ and SU(3) symmetries are coloured in black, red, green and blue, respectively. 

We now consider supersymmetric flows connecting these points for generic $\omega$ in the open range $0 < \omega <\frac{\pi}{8}$. As already emphasised, the masses of the supergravity fields at each critical point do not depend on $\omega$, so the linearised solution (\ref{eq.perturb}) to the BPS equations (\ref{eq.BPSs}) is still controlled by the quantities $\tilde \Delta$ highlighted in bold in table \ref{Table:Masses}. The corresponding AdS$_4$ radii $L_*$ do now depend on $\omega$. As in the case of the previous subsection, the solution also depends on four real integration constants, which control the boundary conditions. 

The pattern of flows between the four critical points  in the right $z$ semidisk (and the corresponding region in the $\zeta_{12}$ disk) of figure \ref{fig.0p01bis} is qualitatively similar to that of the $\omega = 0$ case, although there is also an important difference to be discussed shortly. There exist unique flows from the central SO(8) point in the UV to either G$_2$ point in quadrants 1 and 4 in the IR (preserving G$_2$ symmetry along the flow), and unique flows from either of these G$_2$ points in the UV towards the SU(3)$\times$U(1) point in the IR (preserving SU(3) symmetry along the flow). Additionally, there exists a one-parametric cone of flows between the SO(8) point in the UV to the SU(3)$\times$U(1) point in the IR, bounded by the previous flows. One flow in the family is special, in that it is a {\it direct} flow that minimises the distance in field space between the SO(8) UV and the SU(3)$\times$U(1) IR, and is driven only by an operator of dimension $\Delta_+ = 4.562$ in the IR. The supersymmetry of the direct flow is presumably enhanced to $\cN=2$ in the full $\cN=8$ theory, but we have not explicitly checked this. It is interesting to observe that the trajectory of this direct flow towards the SU(3)$\times$U(1) point in quadrant 1 (and the analog direct flow towards the new SU(3)$\times$U(1) point in quadrant 3) is along the axes of the $\zeta_{12}$ disk and along a curve in the $z$ disk given analytically by
\begin{equation} \label{eq:SU3U1traj}
(1+z \bar z) \big(z e^{-2i\omega} - \bar z e^{2i\omega} \big) + z^2 -\bar z^2 =0 \, ,
\end{equation}
for all $\omega$ including the endpoints of its allowed interval. At $\omega = 0$, the direct flow follows the real $z$ axis (see figure \ref{fig.wis0}), which indeed solves (\ref{eq:SU3U1traj}),  while for $\omega = \frac{\pi}{8}$, this equation corresponds to the red curve in figure \ref{fig.piover8}. We find it rather intriguing that this particular equation turns out to govern the trajectory of the direct flow towards the $\cN=2$ SU(3)$\times$U(1) points, as it already appears in other, different though related, context. Indeed, equation (\ref{eq:SU3U1traj}) appears in one of the conditional clauses (see equation (C.4) of \cite{Borghese:2012zs}),
\begin{eqnarray} \label{N=2cond3}
&& \textrm{either} 
\quad \bar\zeta_1 \zeta_2  =0 \qquad
 \textrm{or} \qquad 
(1+z \bar z) \big(z e^{-2i\omega} - \bar z e^{2i\omega} \big) + z^2 -\bar z^2 =0 \, ,
\end{eqnarray}
that special geometry imposes \cite{Hristov:2009uj,Louis:2012ux} on a critical point to preserve the full $\cN=2$ supersymmetry of the SU(3)-invariant sector action (\ref{electricN=2action}). It is thus curious that an equation that in principle should only know about $\cN=2$ {\it points}, turns out to also know about {\it flows} between $\cN=2$ points\footnote{ The $\cN=8$ SO(8) point is also $\cN=2$ within the truncation corresponding to the SU(3)-singlet sector. Incidentally, such flows can only by $\cN=1$ when considered within the truncated $\cN=2$ theory, as already remarked in footnote \ref{ftnt:N=1inN=2}.}. Even more mysterious is the fact that, although it does, equation (\ref{eq:SU3U1traj}) does not even need to hold at either SO(8) or SU(3)$\times$U(1) points, as the conditional clause (\ref{N=2cond3}) is already fulfilled by $\bar\zeta_1 \zeta_2  =0$ at both $\cN=2$ endpoints (and, in fact, along the entire flow).

The (large $N$) dual field theory interpretation of these domain walls within the right $z$ semidisk (and the corresponding region of the $\zeta_{12}$ disk), in terms of insertions of relevant (irrelevant) operators in the UV (IR) is exactly as in the $\omega = 0$ case \cite{Bobev:2009ms} that we reviewed in section \ref{w=0Flows}, by virtue of the $\omega$-independence of the mass spectra. The only qualitative difference with respect to the case $\omega = 0$ is that now both G$_2$ points are physically inequivalent, and thus describe distinct, G$_2$-symmetric superconformal phases of the dual field theory. The domain walls ending at either G$_2$ point in the UV or the IR describe, for the same reason, inequivalent RG flows of the dual, large $N$ field theory. In order to avoid overloading the paper with figures, we have only plotted, in figure \ref{fig.0p01bis}, the numerical trajectories of two of the latter flows in the cone, that follow a path very close to that of the boundary, and pass near the G$_2$ points without ever reaching them. Physically independent flows now fill out the entire red lobe in the right $z$ semidisk in figure \ref{fig.0p01bis} (and their counterpart in the $\zeta_{12}$ plane), unlike in the $\omega=0$ case, where they only fill half the red lobe of figure \ref{fig.wis0bis}.

The situation changes dramatically on the left $z$ semidisk. Two new one-parametric cones of flows from the SO(8) UV point emerge. They are respectively dominated in the IR by the new SU(3)$\times$U(1) point and by the new SU(3) point. There exist also a unique flow from the SO(8) UV to the new G$_2$ point in the IR, and two unique different flows from the latter in the UV to either the SU(3) or the SU(3)$\times$U(1) points in the IR. Further, the latter two IR points can be also reached by unique RG flows whose UV origin lie in one of the G$_2$ points on the right $z$ semidisk. Finally, in each one of the two new cones there exists a preferred flow, which minimises the distance between the SO(8) UV point and either the SU(3) or the SU(3)$\times$U(1) point in the IR. While we expect the former direct flow to still preserve $\cN=1$ supersymmetry and SU(3) bosonic symmetry as all other generic flows, we expect the supersymmetry and bosonic symmetry of the latter to be enhanced to $\cN=2$ and SU(3)$\times$U(1) in the full $\cN=8$ supergravity. 

In figure \ref{fig.0p01bis} we have again omitted the new direct and unique flows, and have only plotted the trajectories of some generic flows in the new cones. The numerically generated dashed blue trajectories correspond to two different flows in the SU(3)-dominated cone, that leave the SO(8) UV and pass very close to one of the distinct G$_2$ points in quadrants 1 and 3. Similarly, the dashed red curves correspond to different flows in the new SU(3)$\times$U(1) cone that follow paths very close to the boundaries of the cone. The holographic interpretation of the new cones of flows is similar to that of the $\omega =0$ case, again with the difference that the two boundary G$_2$ points and walls correspond to different physical phases and flows. 

In summary, three one-parametric cones of flows emerge for $\omega$ in the open range $0< \omega < \frac{\pi}{8}$, with UV origin in the $\cN=8$ SO(8) point and respectively controlled in the IR by the two distinct $\cN=2$ SU(3)$\times$U(1) points and the $\cN=1$ SU(3) point. The boundaries of these cones correspond to domain walls linking the different G$_2$ points, either in the IR or in the UV, with all other extrema. For generic $\omega$, the pattern of flows is thus a richer version of that in the $\omega =0$ case, and differs from this case in the structure of the cone boundaries. All three lobes in figure \ref{fig.0p01bis} are physically different to one another, and are filled out by physically distinct flows. The symmetry of the $\omega =0$ pattern of flows that figures 
\ref{fig.wis0} and \ref{fig.wis0bis} clearly display is lost when $\omega$ is turned on (compare with figure \ref{fig.0p01bis}). A symmetric pattern is only recovered when $\omega$ reaches the other end of its allowed interval, as we will show next.

\subsection{Flows at $\omega = \frac{\pi}{8}$ }  \label{Flowsw=ToPion8}

\begin{figure}[tb]
\begin{center}
\includegraphics[scale=0.45]{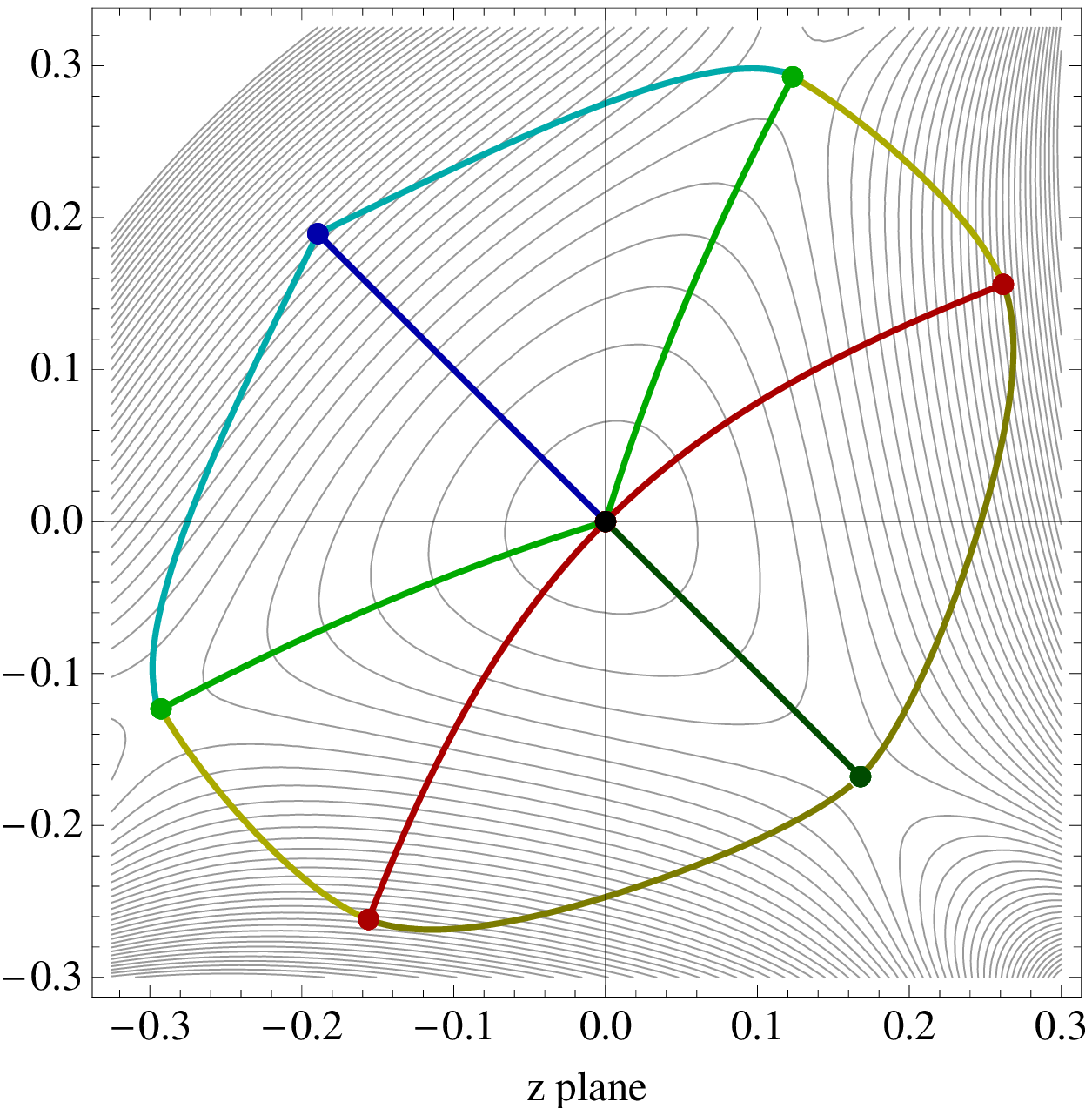}
\includegraphics[scale=0.45]{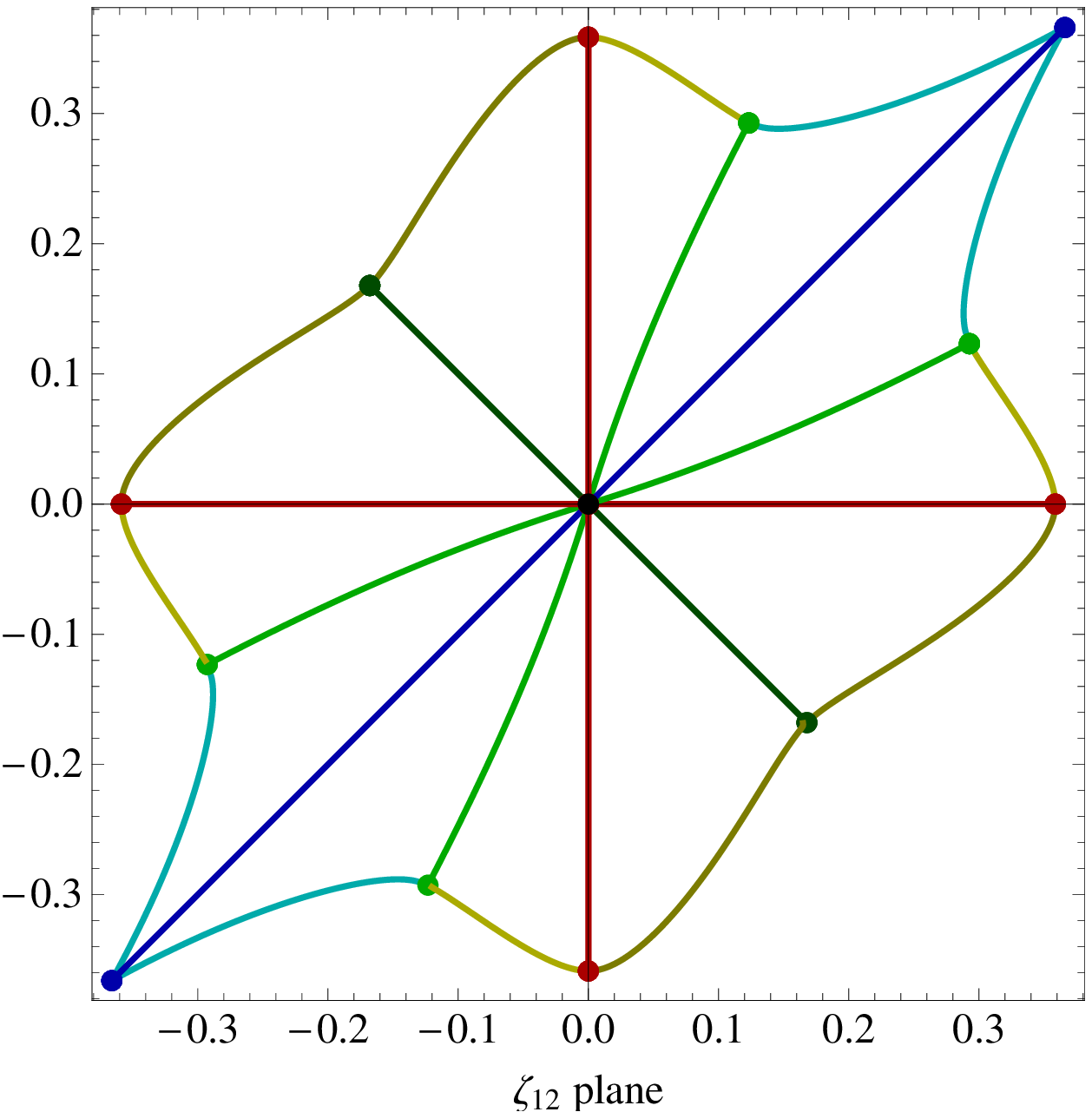}
\caption{The trajectories of the $\omega = \frac{\pi}{8}$ unique and direct flows. Colour code as in previous plots: the SO(8), G$_2$, SU(3)$\times$U(1) and SU(3) points are shown as black, green, red and blue dots, respectively. Points and flows symmetric under a reflection about the NW-SE diagonal are physically equivalent. Different colour shades correspond to different physically distinct points and flows. }\label{fig.piover8}
\end{center}
\end{figure}

If we let $\omega$ evolve further, the seven supersymmetric critical points of the SU(3)-invariant sector continue their migration in the $z$ and $\zeta_{12}$ disks, varying also their associated cosmological constants. When the rightmost end, $\omega = \frac{\pi}{8}$, of the physically allowed range of the electric/magnetic duality angle is reached, the position of the critical points in the disks turns out to be symmetric again, like the case $\omega = 0$ and unlike the case $0<\omega <\frac{\pi}{8}$, although with a different symmetry pattern than when $\omega = 0$ as we will now discuss. When  $\omega = \frac{\pi}{8}$, the seven geometric critical points lie at \cite{Borghese:2012zs}
\begin{align}
\textrm{SO(8)} \rightarrow & \quad z = \zeta_{12} = 0 \label{eq:SO8pointpion8} \ , \\
\textrm{G}_2 \rightarrow & \begin{cases}  z= \pm \zeta_{12} =& \frac{\sqrt{2}+\sqrt{3}-\sqrt{3+2\sqrt{6}}}{2} (1-i) \ ,  \\
 z = \pm \zeta_{12} \approx & 0.12323+0.29371 \, i \ ,\label{eq.g2} \\
  z  = \pm \zeta_{12} \approx & -0.29371- 0.12323 \, i \ , \end{cases} \\
\textrm{SU(3)}\times \textrm{ U(1) } \rightarrow &  \begin{cases} z \approx 0.26185 +0.15608\, i \ , \quad \zeta_{12} = \pm 0.35865\, i  \ , \label{eq.su3u1}\\
 z \approx - 0.15608 - 0.26185 \, i \ , \quad \zeta_{12} = \pm 0.35865\ ,\end{cases} \\
\textrm{ SU(3) } \rightarrow &  \quad z= \left( \sqrt{\frac{3}{2}} - \sqrt{2} \right)(1-i) \ ,\quad \zeta_{12}= \pm \frac{1-\sqrt{3}}{2}(1+i) \ . \label{eq.su3}
\end{align}
We have represented the location of these points in the $z$ and $\zeta_{12}$ disks in figure \ref{fig.piover8}, with the same colour code as in the previous section.

For $\omega =\frac{\pi}{8}$ as for all other values of $\omega$, the position of the critical points is characterised by a value of $z$ and either one of two values of $\zeta_{12}$. What is special about $\omega =\frac{\pi}{8}$ is that the critical points are arranged in a symmetric fashion around the NW-SE diagonal of both $z$ and $\zeta_{12}$ disks, as is evident from (\ref{eq:SO8pointpion8})--(\ref{eq.su3}) and figure \ref{fig.piover8}. Furthermore, points with the same bosonic symmetry and supersymmetry that are mapped into each other under a $\mathbb{Z}_2$ flip 
\begin{eqnarray} \label{NWSEflip}
z \rightarrow -i \bar z \; , \qquad \zeta_{12} \rightarrow -i \bar \zeta_{12}
\end{eqnarray}
about the NW-SE diagonal also have the same cosmological constants. These $\mathbb{Z}_2$-related points should thus be dual to indistinguishable superconformal phases of the dual large $N$ field theory, and thus describe the same physics. We thus take the two $\cN=2$ SU(3)$\times$U(1) points to be identical. Likewise, the G$_2$ points in quadrants 1 and 3 in the $z$ plane of figure \ref{fig.piover8} are physically equivalent, and different to the G$_2$ point that lies on the NW-SE diagonal of quadrant 4, which we henceforth refer to as the $\bar{\textrm{G}}_2$ point. Interestingly, the evolution in $\omega$ renders inequivalent the points with G$_2$ symmetry that were equivalent at $\omega=0$ and, at $\omega=\frac{\pi}{8}$, one of these points is identified with the third G$_2$ point that lives in the boundary $|z|=1$, $|\zeta_{12}|=1$ at $\omega =0$. As advertised in section \ref{sec:SU3inv}, we are thus eventually left with only five physically independent points at $\omega =\frac{\pi}{8}$. Their associated AdS radii are
\begin{align}
& L_{\textrm{SO(8)}}=1 \ , \quad L_{\textrm{SU(3)}\times \textrm{U(1)}}\simeq 0.847468  \ , \quad L_{\textrm{SU(3)}}=\frac{1}{3^{1/4}} \simeq 0.759836 \ , \\
&  L_{\textrm{G}_2}\simeq 0.869127 \ , \quad L_{\bar{\textrm{G}}_2}=\frac{5^{5/4}}{2} \left( 117+62\sqrt{6}\right)^{-1/4} \simeq 0.923204 \ ,
\end{align}
leading to the values of the cosmological constants listed in table \ref{Table:SU(3)}.

The discussion of supersymmetric domain walls between the fixed points proceeds similarly as in the previous cases. Now, the symmetry of the  $\omega = \frac{\pi}{8}$ case allows us to find some of these flows analytically. The discussion of whether a given critical point dominates the IR of a single unique flow, or rather of a one-parameter family of flows, is again exactly as in the $\omega= 0$ case \cite{Bobev:2009ms} that we reviewed in section \ref{w=0Flows}, so here we will just summarise our new results. There exist unique flows between the central, SO(8) point in the UV toward the G$_2$ and $\bar{\mathrm{G}}_2$ points in the IR, whose trajectories follow the green curves in figure \ref{fig.piover8}. The flow towards $\bar{\mathrm{G}}_2$, in particular, proceeds along the NW-SE diagonal. Introducing a real field $x$ along the diagonal as $z = \zeta_{12} = \frac{1-i}{\sqrt{2}} \ x $, the BPS flow equations (\ref{eq.BPSs}) reduce to a single equation,
\begin{eqnarray}
\frac{dx}{dr} = -  \frac{x (1+x) (x^4-4x^3-4x+1) }{\left( 1-x^2  \right)^{5/2} } \ , 
\end{eqnarray}
which a computer easily integrates into a (rather unenlightening) analytical expression for the inverse radial profile $r=r(x)$ of the field $x$ along to the $\bar{\mathrm{G}}_2$ flow. There are also unique interpolating domain walls connecting the points with G$_2$ symmetry in the UV with the points with SU(3)$\times$U(1) symmetry in the IR, whose numerically generated trajectories follow the  yellow curves in figure \ref{fig.piover8}. Observe that now, like in the $\omega = 0$ case, there exists only one physically distinct $\cN=2$ SU(3)$\times$U(1) point. However, unlike in the case $\omega = 0$, where that point arises as the IR end of RG flows originating in two physically equivalent UV G$_2$ points, now the $\cN=2$ point can be reached from two physically different UV phases G$_2$ and $\bar{\textrm{G}}_2$. The list of unique flows closes with those interpolating from the two physically equivalent G$_2$ points in the UV and the SU(3) phase in the IR. We have also generated numerically the radial profiles of both physically equivalent flows, and have depicted (in cyan) their resulting trajectories in figure \ref{fig.piover8}.

\begin{figure}[tb]
\begin{center}
\includegraphics[scale=0.45]{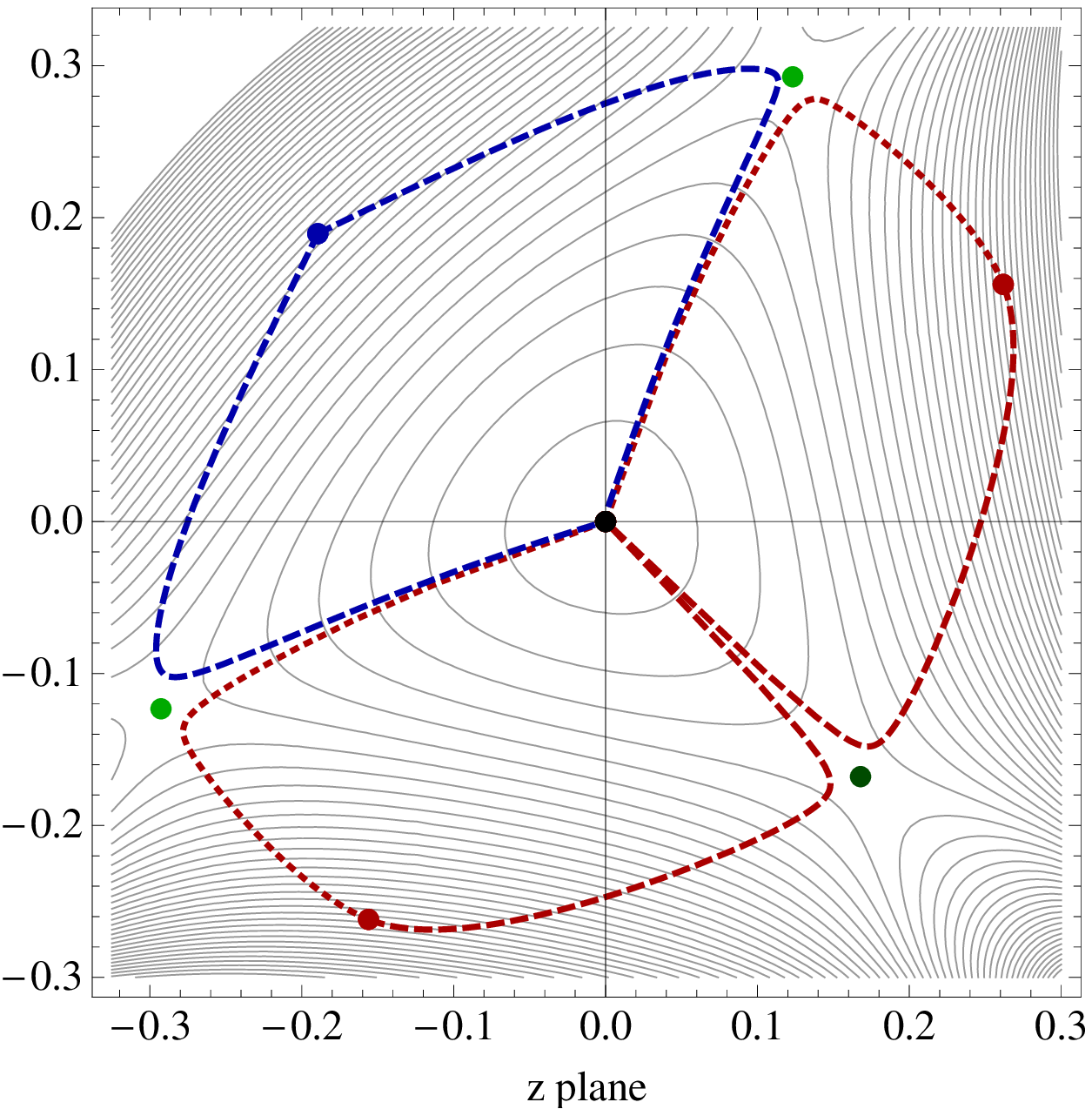}
\includegraphics[scale=0.45]{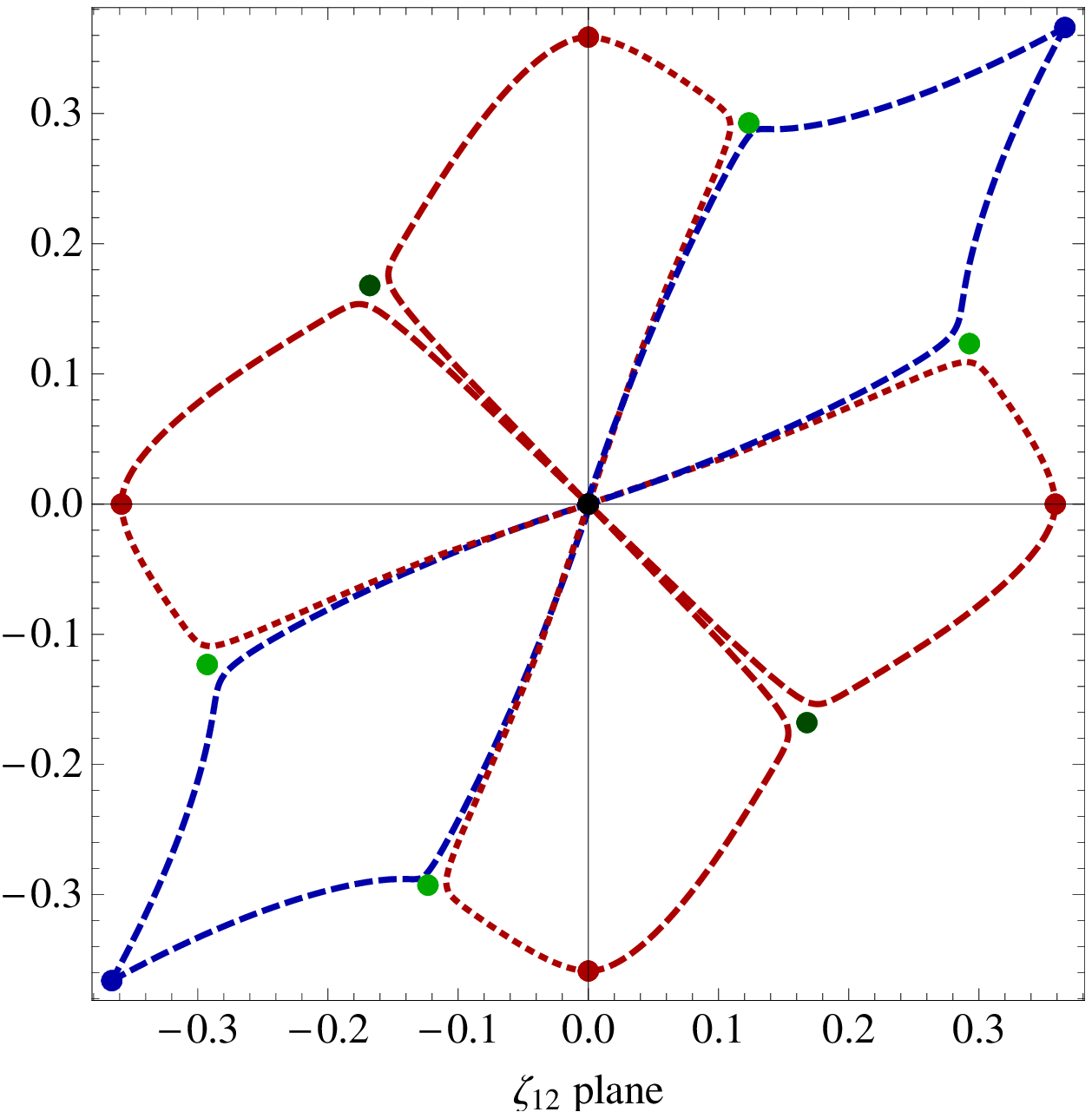}
\caption{Some flows at $\omega=\frac{\pi}{8}$. Colour code as in previous plots. The plotted flows follow paths very close to the boundaries of the two physically inequivalent cones (the two red lobes in the $z$ disk, and their $\zeta_{12}$ companions, are identified). Different colour shades and different dashings correspond to different physically distinct points and flows.}\label{fig.piover8bis}
\end{center}
\end{figure}

More generally, there exist one-parameter cones of flows with SO(8) UV origin and IR end in the SU(3)$\times$U(1) and SU(3) points. Note that the $\mathbb{Z}_2$ symmetry (\ref{NWSEflip}) now identifies the two cones with SU(3)$\times$U(1) in the IR of the generic $\omega$ case of section \ref{wIn0ToPion8}, into just one physically meaningful cone, as in the $\omega =0$ case. Unlike $\omega =0$, however, where the SU(3)$\times$U(1) IR-dominated cone is half-filled so that it is bounded by flows through G$_2$ and the direct one between SO(8) to SU(3)$\times$U(1), its $\omega = \frac{\pi}{8}$ counterpart is filled out completely and is only bounded by G$_2$ flows. Additionally, we find a one-parameter cone of flows with IR governed by the $\cN=1$ SU(3) point, which does not have an analog in the $\omega =0$ theory. We have again verified that, within each cone of flows, there exist preferred flows between the SO(8) UV and the corresponding IR point, which are direct in the sense that minimise the distance in field space between the endpoints of the flow. On the one hand, the direct flow towards the $\cN=1$ SU(3) point proceeds along the NW-SE diagonal of quadrant 2 in the $z$ plane, and its path corresponds to the straight blue line in figure \ref{fig.piover8}. The integration of this direct flow is simplified by introducing real scalars $x$ and $y$ along the relevant diagonals of the $z$ and $\zeta_{12}$ planes as $z = \frac{i-1}{\sqrt{2}} \ x $ and $\zeta_{12} = \frac{i+1}{\sqrt{2}} \ y $. The BPS equations (\ref{eq.BPSs})  then reduce to a simpler set of equations for $x$ and $y$ only. Due to the $\mathbb{Z}_2$ identification, this direct flow serves as a boundary, together with the corresponding G$_2$ flows, of the SU(3) cone. For the other direct flow towards the SU(3)$\times$U(1) point,  on the other hand, we have verified that it is driven toward the infrarred by only one mode that falls-off with  $\tilde \Delta = \Delta_- = -1.562$ in equation (\ref{eq.perturb}), like its analog $\omega = 0$ flow \cite{Ahn:2000aq}. We have again generated numerically the radial profiles for this direct flow. The trajectories in field space of the two physically equivalent SU(3)$\times$U(1) direct flows are the red curves in figure 
\ref{fig.piover8}. As already mentioned in section \ref{wIn0ToPion8}, these curves turn out be analytically given by equation (\ref{eq:SU3U1traj}).

As in the previous cases, a generic flow within each cone follows a steepest descent path of the superpotential that does not minimise the trajectory between the UV and IR phases in field space. Some flows within the cones follow trajectories very close to the boundary flows governed by the points with G$_2$ symmetry, and approach arbitrarily close to those points before following their descent into the appropriate, SU(3) or SU(3)$\times$U(1), IR fixed points. 
In figure \ref{fig.piover8bis} we present some examples of these types of flows.

Let us discuss in some detail the features of the holography of the new cone of domain walls dominated by the $\cN=1$ SU(3) point in the IR. From table \ref{Table:Masses}, it can be seen that a generic flow into the SU(3) point is driven by a combination of two deformations (corresponding to non-normalisable fall-offs $\tilde \Delta = \Delta_- = -1.450$) of the dual IR field theory lagrangian by irrelevant operators of dimension $\Delta = \Delta_+ = 4.450$. Both modes are active in the unique, boundary flow from the G$_2$ point in the UV, where the flow is caused (as in the $\omega =0$ flow to SU(3)$\times$U(1)) by a deformation of the G$_2$-symmetric UV phase by relevant deformations of dimension $\Delta_+ = 1.592$ and $\Delta_+ = 2.408$. The other boundary of the cone corresponds to the direct flow from SO(8). The latter is caused by a mass deformation that drives the flow into the SU(3) point, where it lands driven by the insertion of only one of the irrelevant operators of dimension $\Delta = \Delta_+ = 4.450$.

\begin{figure}[tb]
\begin{center}
\includegraphics[scale=0.5]{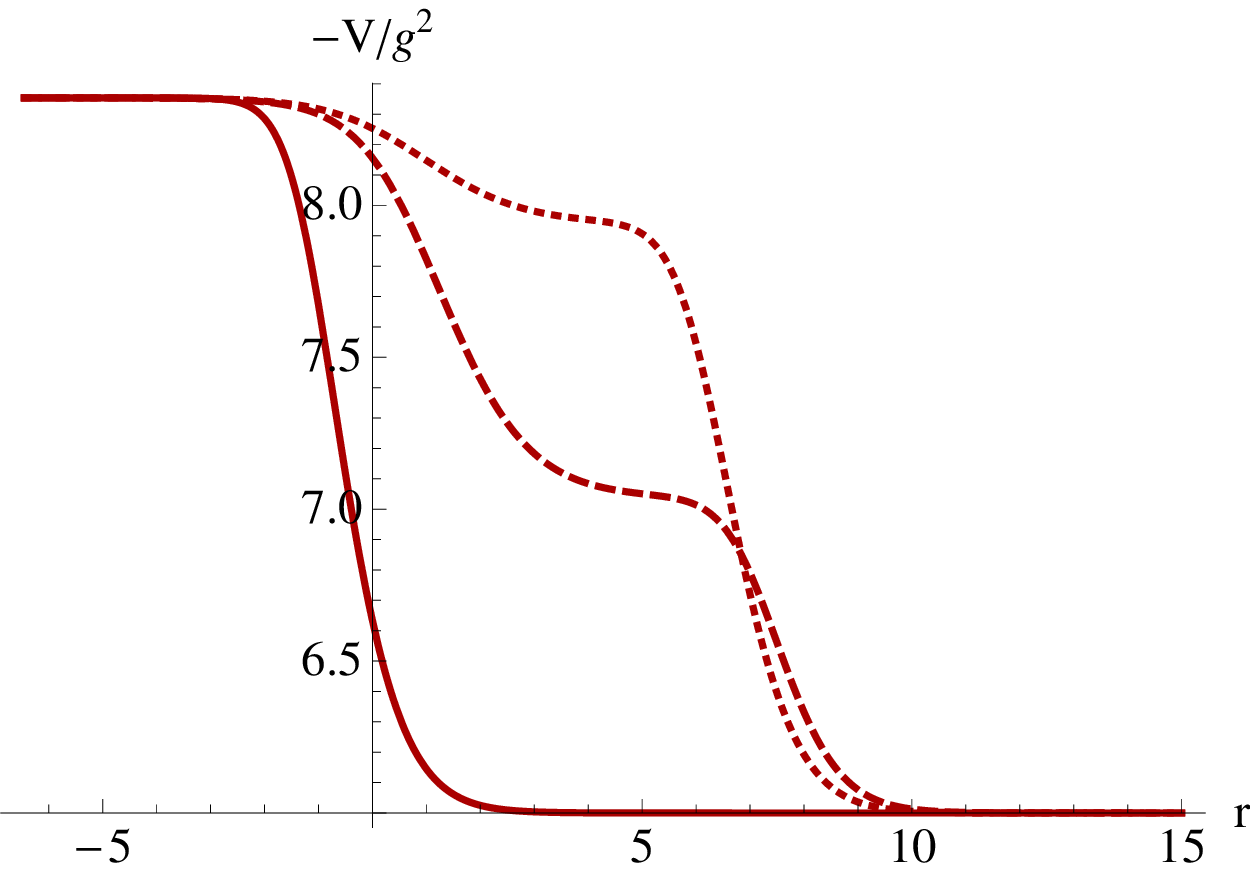}
\includegraphics[scale=0.5]{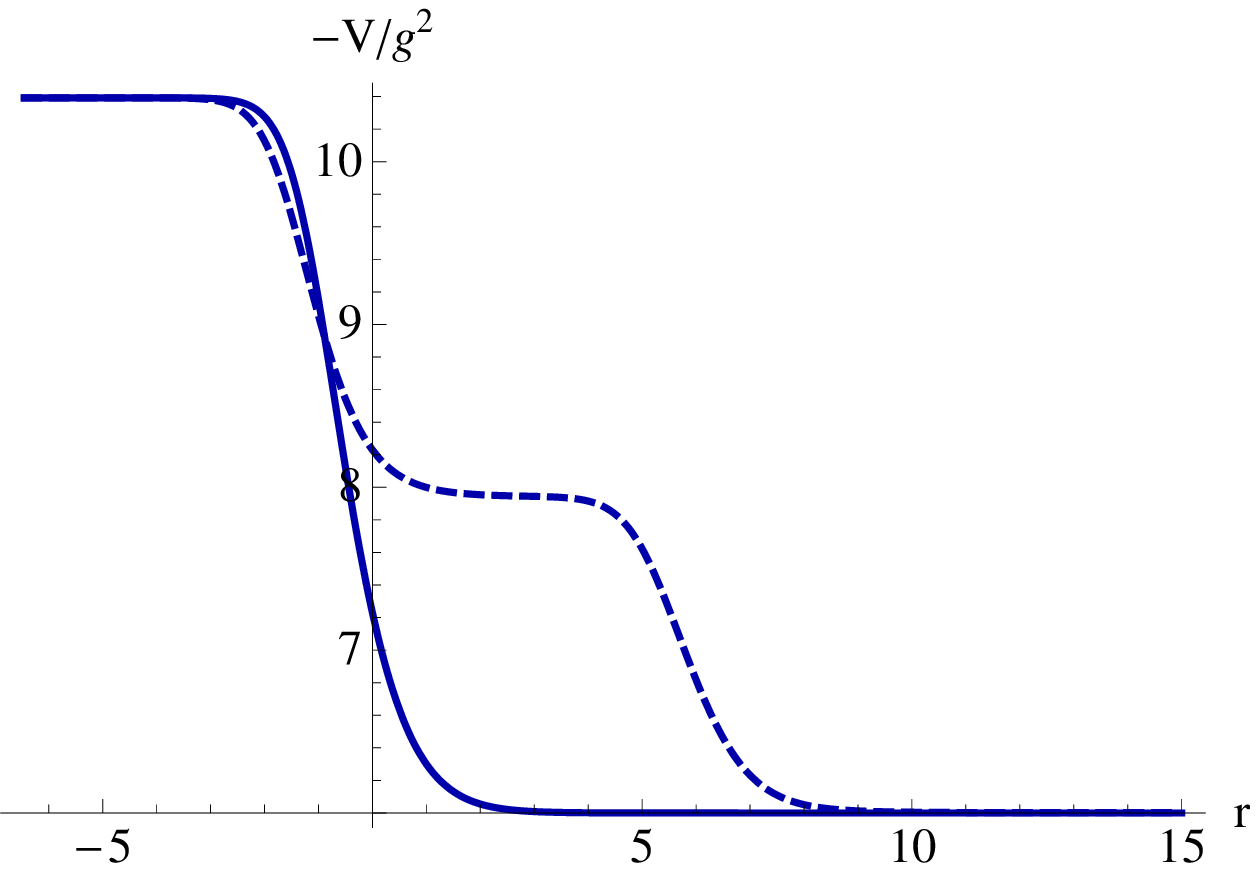}
\caption{Left: the value of (minus) the potential \eqref{PotFromSuperPot} along the two flows in the $\omega= \frac{\pi}{8}$ SU(3)$\times$U(1) cone that pass close to the G$_2$ and $\bar{\textrm{G}}_2$ fixed points (dotted and dashed lines) in figure \ref{fig.piover8bis} and the direct flow (continuous line) in figure \ref{fig.piover8}. Right: the same for the flows in the $\omega= \frac{\pi}{8}$ SU(3) cone depicted in figures \ref{fig.piover8bis}  (dashed) and \ref{fig.piover8} (continuous lines). The plateaux roughly correspond to the cosmological constant at the G$_2$ or $\bar{\textrm{G}}_2$ points (see table \ref{Table:SU(3)}).}\label{fig.potentialpiover8}
\end{center}
\end{figure}

It is also interesting to study the behaviour of the scalar potential (\ref{PotFromSuperPot}) along generic domain walls in each cone (see figure \ref{fig.potentialpiover8}). A plot of the (negative of the) potential as a function of the radial coordinate for some of the flows in each cone shows a monotonically decreasing function that asymptotes to the values of the potential at the UV and IR fixed points. This is consistent with the fact that the flows proceed from higher (in the $r \rightarrow \infty$ UV) to lower (in the $r \rightarrow -\infty$ IR) values of the scalar potential (with sign). Flows in a given cone different from the direct one  exhibit a plateau governed by the corresponding limiting G$_2$  point, which is shorter or longer depending on whether the flow follows a path closer to the direct flow or to the G$_2$ point. The potential at a plateau roughly matches the scalar potential at the G$_2$ point by which it is governed.
In the left panel of figure \ref{fig.potentialpiover8} we show the value of the cosmological constant along the two distinct $\omega = \frac{\pi}{8}$ flows depicted in figure \ref{fig.piover8bis}   that end in the SU(3)$\times$U(1) fixed point in the first quadrant of the $z$ disk in that figure,  along with the direct flow shown in red in figure \ref{fig.piover8}.  The analog flows to the SU(3)$\times$U(1) point in the third quadrant  of figure  \ref{fig.piover8bis} are physically equivalent to the former two, and accordingly give rise to the same graph for the radial evolution of the cosmological constant, as we have verified. The right panel of figure  \ref{fig.potentialpiover8} corresponds to the potential along the flows depicted with a dashed blue line in figure \ref{fig.piover8bis} and a continuous blue line in figure \ref{fig.piover8}, that are dominated by the SU(3) IR point. Observe that there are two plateaux in the plot corresponding to SU(3)$\times$U(1), whereas there is only one for that corresponding to SU(3). This is due to the fact  that the boundaries of the $\omega =\frac{\pi}{8}$ SU(3)$\times$U(1) cone are dominated by two different, G$_2$ and $\bar{\textrm{G}}_2$ points, whereas the boundary of the $\omega =\frac{\pi}{8}$ SU(3) cone is governed by just a physically distinct G$_2$ point. Observe also that both plots display a plateau around $V = -8.354 g^2$, corresponding to the scalar potential of the G$_2$ point (see table \ref{Table:SU(3)}) common to the boundary of both cones. Finally, we note that the plot of the potential along flows in the $\omega =0$ SU(3)$\times$U(1) cone displays the same qualitative behaviour of the right panel of figure  \ref{fig.potentialpiover8}, only with a plateau roughly at the value $V = -7.192 g^2$ of the cosmological constant of the $\omega=0$ G$_2$ critical point.

To summarise, at $\omega = \frac{\pi}{8}$, we find unique, G$_2$-symmetric flows between the central SO(8) point in the UV and two inequivalent G$_2$ and $\bar{\textrm{G}}_2$ points in the IR. These are respectively linked by SU(3)-symmetric flows to the SU(3) and SU(3)$\times$U(1) points. Furthermore, there exist two cones of flows, dominated by the latter points in the IR and the SO(8) point in the UV. The SU(3)$\times$U(1) cone is bounded by flows with either G$_2$ and $\bar{\textrm{G}}_2$ UV or IR endpoints, while the SU(3) cone is limited by the direct flow from SO(8) and the unique flow from G$_2$. Table \ref{Table:cones} in the introduction provides a summary of all possible cases that arise when $\omega$ is varied.


\section{The Coulomb branch} \label{sec.coulombbranch}

In the previous section, we studied holographic RG flows due to insertions of relevant operators in the UV conformal theory.  We can instead deform the theory with vevs, and study  the Coulomb branch. As equation \eqref{eq.UVperturb2} shows, the generic deformation around the SO(8) UV fixed point is described by  non-normalisable modes ($\tilde\Delta_i=1$) that can be chosen to correspond either to insertion in the lagrangian, as in the previous section, or as vevs. In this section, we initiate the study of the latter type of flows for the theories in  \cite{Dall'Agata:2012bb}.

\subsection{Truncation to one scalar}

Following \cite{Freedman:1999gk, Cvetic:1999xx}, we now consider scalars in the SL$(8,\mathbb{R})/$SO(8) sector of the E$_{7(7)}/$SU(8) scalar manifold of the full $\cN=8$ theory \cite{Dall'Agata:2012bb}. The superpotential in this sector was given in appendix A of \cite{Borghese:2012zs}. The corresponding theory can be truncated to the sector containing the seven dilatons only, which in turn can be further truncated into seven one-scalar models, $n=1, \ldots, 7$, with SO$(n)\times$SO$(7-n)$ symmetry. Using, for convenience, a different normalisation for the Einstein-Hilbert term from the previous section, we find the lagrangians of the $n=1, \ldots, 7$ models to be
\begin{eqnarray}\label{eq.onefieldmodel}
e^{-1} {\cal L} = R -\tfrac{1}{2} ( \partial \varphi)^2 -V_{n}(\varphi) \ ,
\end{eqnarray}
where the scalar potential $V_{n}$ depends on the scalar $\varphi$, on the phase $\omega$ and on the integer $n=1,\ldots , 7$. Defining
\begin{eqnarray} \label{defXphi}
X_n (\varphi) = e^{\frac{4}{\sqrt{n(8-n)}}\varphi} \ ,
\end{eqnarray}
and the functions
\begin{eqnarray} \label{potdefpieces}
& \tilde{V}_1(x) = \tfrac{1}{4} \big( x^{-7/4} - 14 x^{-3/4} - 35 x^{1/4} \big) \, , \quad &
\tilde{V}_2(x) = -6 \big( x^{-1/2} + x^{1/2}  \big) \, , \nonumber \\
& \tilde{V}_3(x) = -\tfrac{3}{4}   \big( x^{-5/4} + 10 x^{-1/4} + 5 x^{3/4} \big) \, , \quad &
\tilde{V}_4 (x) = -2 \big( x^{-1} + x + 4   \big) \, , 
\end{eqnarray}
the scalar potential for each value of $n$ is defined as a function of $\varphi$ through $X_n(\varphi)$ as
\begin{eqnarray} \label{potdef}
&& V_n (\varphi)= g^2 \, \tilde{V}_n (X_n) \cos^2 \omega + g^2  \, \tilde{V}_{n} (X_n^{-1}) \sin^2 \omega \ ,\qquad \text{for }n=1,2,3,4 \ , \\
&& V_n (\varphi)= g^2 \, \tilde{V}_{8-n} (X_n^{-1}) \cos^2 \omega + g^2 \, \tilde{V}_{8-n} (X_n) \sin^2 \omega \ , \qquad \text{for }n=5,6,7 \ .
\end{eqnarray}
Here, we have reinserted the coupling $g$. It is interesting to note that the $\omega$-dependence dissappears for even values of $n$. 

At fixed $n$, the theory (\ref{eq.onefieldmodel}) can be mapped into itself or into the $8-n$ theory by the discrete symmetries
\begin{align} \label{eq:DiscreteTrans12}
& \left( n, \omega, \varphi \right)  \to \left( n, \tfrac{\pi}{2} - \omega, -\varphi \right) \ , \quad 
 \left( n, \omega, \varphi \right)  \to \left( 8-n, \omega, -\varphi \right)  \ .
\end{align}
These can be combined into a third symmetry
\begin{equation}\label{eq.varphisymmetry}
\left( n, \omega, \varphi \right)  \to \left( 8-n, \tfrac{\pi}{2} - \omega, \varphi \right)  \  ,
\end{equation}
which in particular implies  $V_2(\varphi) = V_6(\varphi)$.

These symmetries allow us to study the system in the range $\omega\in [0,\tfrac{\pi}{4}]$. In the one-scalar truncation (\ref{eq.onefieldmodel}), it is not possible to further reduce the allowed range for the angle $\omega$ to $[0,\tfrac{\pi}{8}]$, as in  the full theory \cite{Dall'Agata:2012bb} and its SU(3)-invariant sector, because some of the modes involved in identifications needed to reduce the periodicity to $\tfrac{\pi}{8}$ have been truncated out. We will focus here on solutions where the IR (which we now set at $r=0$) divergence of the scalar is of the form $\varphi (r \rightarrow 0) \rightarrow -\infty$. Different solutions for which the scalar diverges as $\varphi (r \rightarrow 0) \rightarrow +\infty$ can be studied with the help of the transformations (\ref{eq:DiscreteTrans12}), (\ref{eq.varphisymmetry}).

Finally, observe that the $n=1,2,6,7$ models (\ref{eq.onefieldmodel}) overlap with the SU(3) invariant sector since, for $n=1,7$ the model  (\ref{eq.onefieldmodel}) displays SO(7)$\supset $SU(3) symmetry when embedded into the full $\cN=8$ theory, and for  $n=2,4$, (\ref{eq.onefieldmodel}) has SO(6)$\times$ SO(2)$\supset$ SU(3) symmetry. Table \ref{tab.dilatons} shows the relation between the real dilaton that appears in the model \eqref{eq.onefieldmodel} and the two complex scalars that we used in section \ref{sec.flows} to parametrise the SU(3)-singlet sector. The only supersymmetric extremum of the models  (\ref{eq.onefieldmodel}) occurs at $\varphi=0$, corresponding to the SO(8) critical point with AdS radius $L = -(\sqrt{2}g)^{-1}$.

\begin{table}[tb]
\begin{center}
\begin{tabular}{c|cc}
$n$ & $z$ & $\zeta_{12}$ \\
\hline
$1$ & $\tanh \frac{\varphi}{2\sqrt{7}}$ & $\pm \tanh \frac{\varphi}{2\sqrt{7}}$ \\
$2$ & $\tanh \frac{\varphi}{2\sqrt{3}}$ & $0$ \\
$3,4,5$ & $-$ & $-$ \\
$6$ & $\tanh \frac{-\varphi}{2\sqrt{3}}$ & $0$ \\
$7$ & $\tanh \frac{-\varphi}{2\sqrt{7}}$ & $\pm \tanh \frac{-\varphi}{2\sqrt{7}}$  
\end{tabular}
\caption{Relations between the real dilaton in the one-scalar truncation (\ref{eq.onefieldmodel}) and the two complex scalars of the SU(3)-invariant sector of section \ref{sec.flows}.} \label{tab.dilatons}
\end{center}
\end{table}

\subsection{Coulomb branch at $\omega=0$} \label{sec:Coulombw=0}

Let us first review the $\omega=0$ case, studied in \cite{Cvetic:1999xx}, in order to fix our conventions and gear up for the $\omega \neq 0$ case of next section. 

The equations of motion that derive  from \eqref{eq.onefieldmodel} can be expressed in terms of first order differential equations. Indeed, introducing the superpotential
\begin{equation}\label{eq.fakesuperpotential}
W_n(\varphi) = \frac{1}{8} e^\frac{n\, \varphi}{2\sqrt{n(8-n)} } \left( 8-n+n \, e^\frac{ - 4 \varphi}{\sqrt{n(8-n)}} \right) \ ,
\end{equation}
from where the $\omega=0$ potential derives as
\begin{equation}
V_n(\varphi) = 4g^2  \left( 4 \left(\partial_\varphi W_n\right)^2 - 3 W_n^2 \right) \ ,
\end{equation}
the second order differential equations that derive from \eqref{eq.onefieldmodel}, evaluated on the domain wall metric \eqref{eq:DW}, are solved by the first order system
\begin{equation}\label{eq.firstorderw0}
\varphi'(r) =  4\sqrt{2} \, g \frac{\partial W_n}{\partial \varphi} \ , \qquad A'(r) = - \sqrt{2}\, g W_n \ .
\end{equation}

\begin{figure}[tb]
\begin{center}
\includegraphics[scale=0.6]{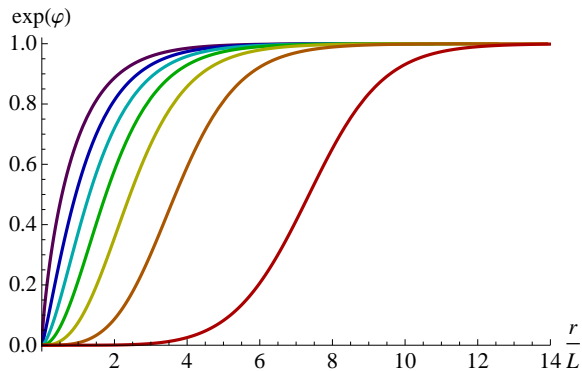}
\caption{Radial profile of the scalar from $n=1$ (leftmost curve) to $n=7$ (rightmost one).}\label{fig.scalars}
\end{center}
\end{figure}

From \eqref{eq.firstorderw0} and \eqref{eq.fakesuperpotential} it is straightforward to express $A$ as a function of the scalar
\begin{equation}\label{eq.Aprofile}
\exp \left( 2 A(\varphi) \right) = e^{-2\left( \gamma_E+\psi \left(1-\frac{n}{8} \right)\right)}e^\frac{n \varphi}{\sqrt{n(8-n)}}   \left( e^\frac{4 \varphi}{\sqrt{n(8-n)}} - 1 \right)^{-2} \ ,
\end{equation}
where  the normalisation is given in terms of the Euler constant,  $\gamma_E$, and  the digamma function, $\psi(x)$. With this choice of integration constant,  the  AdS behaviour, $\exp(2A)= e^{2r/L}$, is recovered asymptotically in the UV ($r \rightarrow \infty$). 
The radial profile for the scalar function, $\varphi(r)$, is obtained via the inverse of the relation
\begin{equation} \label{eq.rvarphi}
r(\varphi) = \frac{8}{8-n} e^\frac{\sqrt{8-n}\,\varphi}{2\sqrt{n}} {}_2F_1\left( 1 , 1- \frac{n}{8} ; 2-\frac{n}{8} ; e^{\frac{4\varphi}{\sqrt{n(8-n)}}} \right) \ .
\end{equation}
Here we have fixed an integration constant in such a way that the scalar diverges at the IR, radial origin,  $\varphi(0)=-\infty$. The profile of the scalars for the seven cases $n=1,\ldots,7$ is shown in figure \ref{fig.scalars}. Specifically, for $n=4$, one obtains the simple analytic result
\begin{equation}
\exp(\varphi)_{n=4} =  \tanh^2 \frac{g\,r}{\sqrt{2}} \quad   \Rightarrow \quad \exp \left( 2 A(\varphi) \right)_{n=4} = 4 \sinh^2 ( \sqrt{2} g r )\ .
\end{equation}

\begin{figure}[tb]
\begin{center}
\includegraphics[scale=0.55]{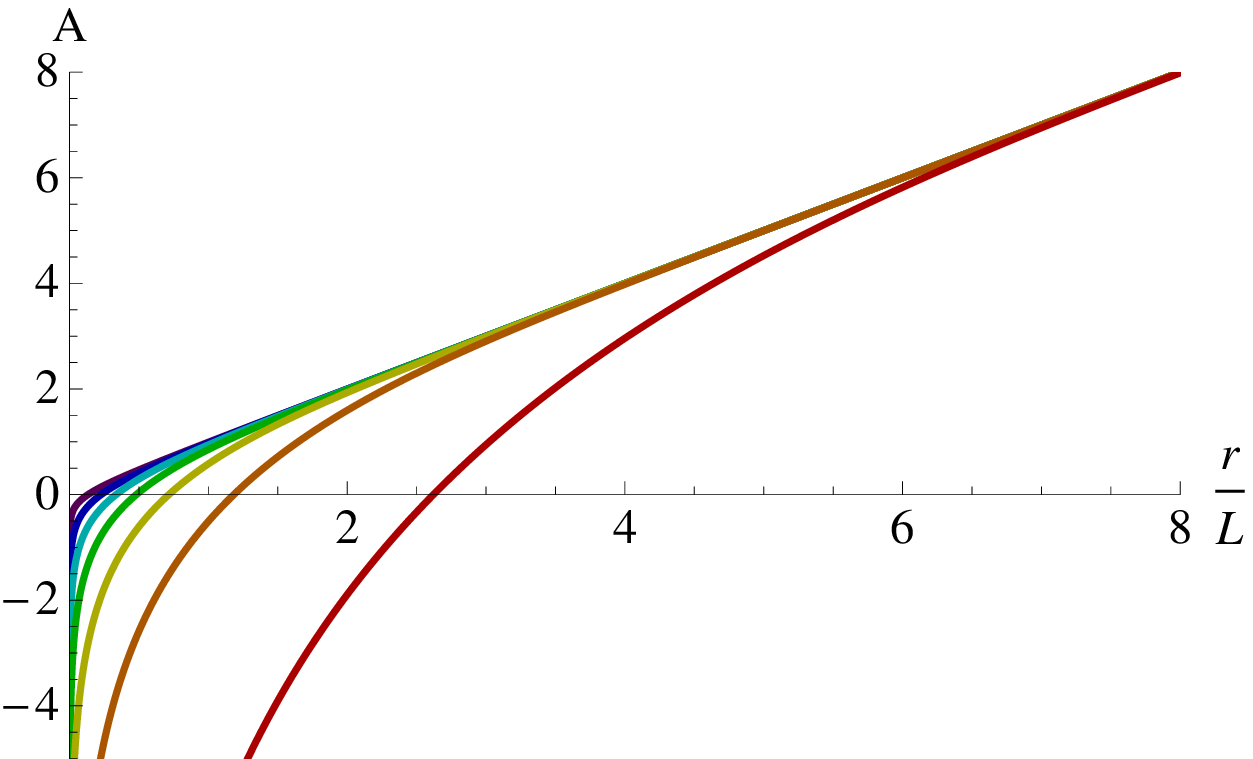}
\includegraphics[scale=0.55]{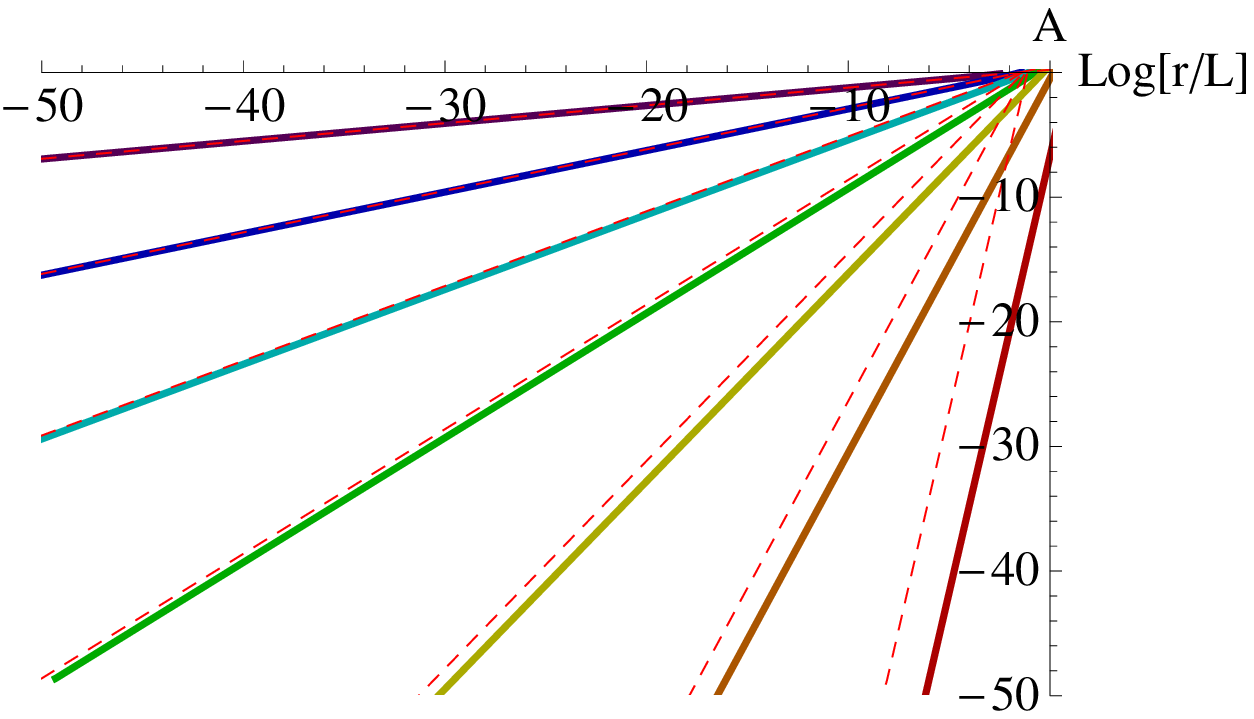}
\caption{The metric function profile $A_n(r)$ for $n=1$ (leftmost curve) through $n=7$ (rightmost one). As is clearly visible in the left plot, all profiles tend to the AdS$_4$ value $A= r/L$ in the UV (large $r$). The right plot zooms into the IR ($r \approx 0$) region, where each fuction approaches the low-$r$ approximation $A=\gamma_E + \psi \left(1-\frac{n}{8} \right) + \frac{n}{8-n} \log[r/L]$, shown as dashed lines.}\label{fig.Aprofiles}
\end{center}
\end{figure}

Given the solution \eqref{eq.rvarphi}, the radial profile of the domain wall function, $A_n(r)$ in (\ref{eq:DW}), is straightforward to calculate for each $n=1, \ldots , 7$. We have numerically plotted it  in figure \ref{fig.Aprofiles}. The near-origin behaviour, where the scalar diverges, is shown explicitly, and compared to the analytic behaviour obtained from \eqref{eq.Aprofile}, $e^{2A} \sim \left(\frac{r}{L}\right)^\frac{2n}{8-n}$. 

A study of the spectrum of a marginal scalar operator on these solutions was performed in \cite{Cvetic:1999xx}. An analysis of the effective potential in the Klein-Gordon equation for a probe massless scalar in these backgrounds reveals that, for $n>4$, there is a continuous spectrum with no mass gap. In the $n=4$ case the spectrum is still continuous, but now a gap appears. Finally, when $n<4$ the spectrum is discrete.


\subsection{Coulomb branch at  $\omega\neq 0$}

We now turn to the Coulomb branch flows for non-zero values of the dyonically-gauging parameter $\omega$, and chose to work in the second order formalism provided by the equations of motion that derive from the lagrangian (\ref{eq.onefieldmodel}). For simplicity we will discuss only the $n=1$ case (and the $n=7$ case, by virtue of the symmetry \eqref{eq.varphisymmetry}).

The scalar potential $V_1(\varphi)$ of the $n=1$ model (\ref{eq.onefieldmodel})  has one non-supersymmetric critical point with residual symmetry SO(7), besides the supersymmetric  SO(8) fixed point. 
This non-supersymmetric point typically affects the behaviour of a generic flow solution of the second order differential equations. As will now argue, the supersymmetric flow is the unique flow for which the presence of the SO(7) point goes unnoticed. 
The generic flow solution to the second order equations of motion involves four constants of integration: two of them are fixed by imposing the behaviour of the metric and scalar  in the IR, and that the divergence of the scalar field occurs at $r=0$. After these asymptotics are imposed, the other two ($\omega$-dependent) constants of integration, $\beta_1$, $\beta_2$,  appear in the low $r$ expansion
\begin{align}\label{eq.boundaryconditionvarphi}
e^{\tfrac{\varphi}{\sqrt{7}}} & = \frac{7^{2/7}}{2^{6/7}} \left(\cos\omega \right)^{2/7} \left( \frac{r}{L} \right)^{2/7} + \frac{3}{2^{6/7}7^{5/7}} \frac{\beta_2}{\beta_1} \left(\cos\omega \right)^{2/7} \left( \frac{r}{L} \right)^{6/7} + {\cal O}\left( \frac{r}{L} \right)^{10/7} \ , \\
\label{eq.boundaryconditionA}
e^A & = \beta_1  \left( \frac{r}{L} \right)^{1/7} + \beta_2  \left( \frac{r}{L} \right)^{5/7} + {\cal O}\left( \frac{r}{L} \right)^{9/7} \ .
\end{align}
The constant $\beta_1$ can be absorbed in a scaling of the Minkowskian directions of the domain wall, so we  set $\beta_1=1$ without loss of generality.
The surviving constant of integration, $\beta_2$, parametrises a family of flows which contains, for a specific,  $\omega$-dependent value of $\beta_2$, the supersymmetric flow.

In order to gain some insight on how to determine the (unique) supersymmetric member in this family of flows, let us first retrieve the $n=1$, $\omega=0$ case from our current second-order-equations perspective. We have checked that, integrating the second order differential equations  at $\omega=0$ with a particular value of $\beta_2$ ($\beta_2 = 0$), we exactly recover the supersymmetric $n=1$ solution of section \ref{sec:Coulombw=0}. In the supersymmetric solution, the scalar asymptotically approaches its critical value in the UV, $\varphi(r\to\infty)=0^-$, monotonically. This monotonic behaviour of the solution is characteristic of the supersymmetric flow, as it is in fact driven by first order supersymmetric equations that relate $\varphi'$ to the position in field space. Now, if we take $\beta_2<0$  in the $\omega = 0$ second order solutions (\ref{eq.boundaryconditionvarphi}),  (\ref{eq.boundaryconditionA}), the solution to the scalar profile still reaches monotonically $\varphi(r\to\infty)=0^-$. But if we take $\beta_2>0$, this monotonicity is lost. The profile of the scalar field now approaches $\varphi(r\to\infty)=0^+$, taking the value $\varphi=0$ both asymptotically and at some finite value of the radial coordinate. As we have just argued, the supersymmetric flow does not allow this kind of behaviour, so the $\beta_2 > 0$ flows are manifestly non supersymmetric. Thus, the value of the integration constant $\beta_2$ that yields the supersymmetric flow is that for which the transition to non-monotic behaviour occurs. 

Still at $\omega=0$, this value of $\beta_2 $ also leads to a manifestly supersymmetric behaviour of the flow, in that it becomes insensitive to the presence of the non-supersymmetric SO(7) fixed point.
If we lower $\beta_2$, we see a plateau arising in some range of the radial coordinate. The height of this plateau is roughly given by the $\varphi$ location of the non-supersymmetric SO(7) critical point. 
Actually, if $\beta_2$ is sufficiently negative, the radial profile starts at $e^{\phi(0)}=0$ in the IR, never reaches the value of the non-supersymmetric SO(7), and diverges again at some finite radius. This is clearly a non-supersymmetric behaviour, since what this describes is a scalar with a boundary condition such that it is not able to pass the non-supersymmetric potential barrier. On the other hand, at $\beta_2 =0$ the plateau disappears, signaling the supersymmetric flow.

\begin{figure}[tb]
\begin{center}
\includegraphics[scale=0.55]{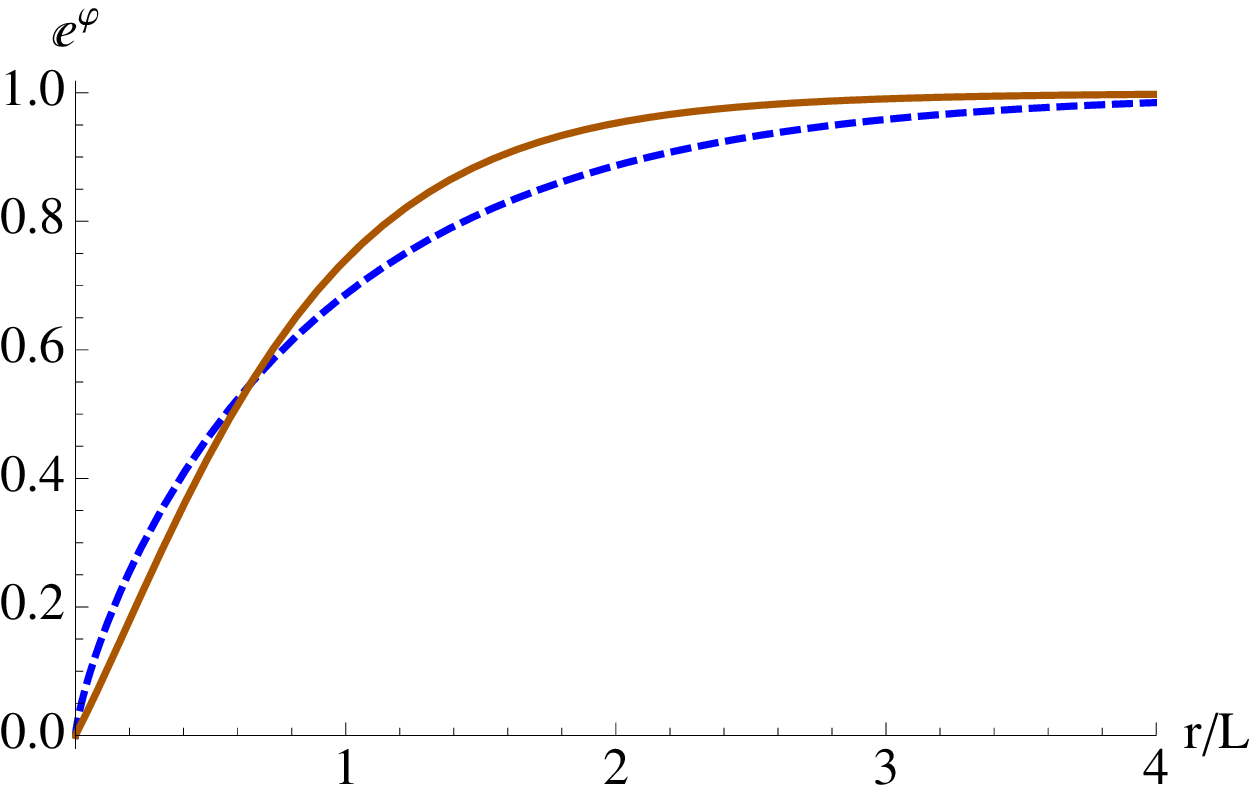}
\includegraphics[scale=0.55]{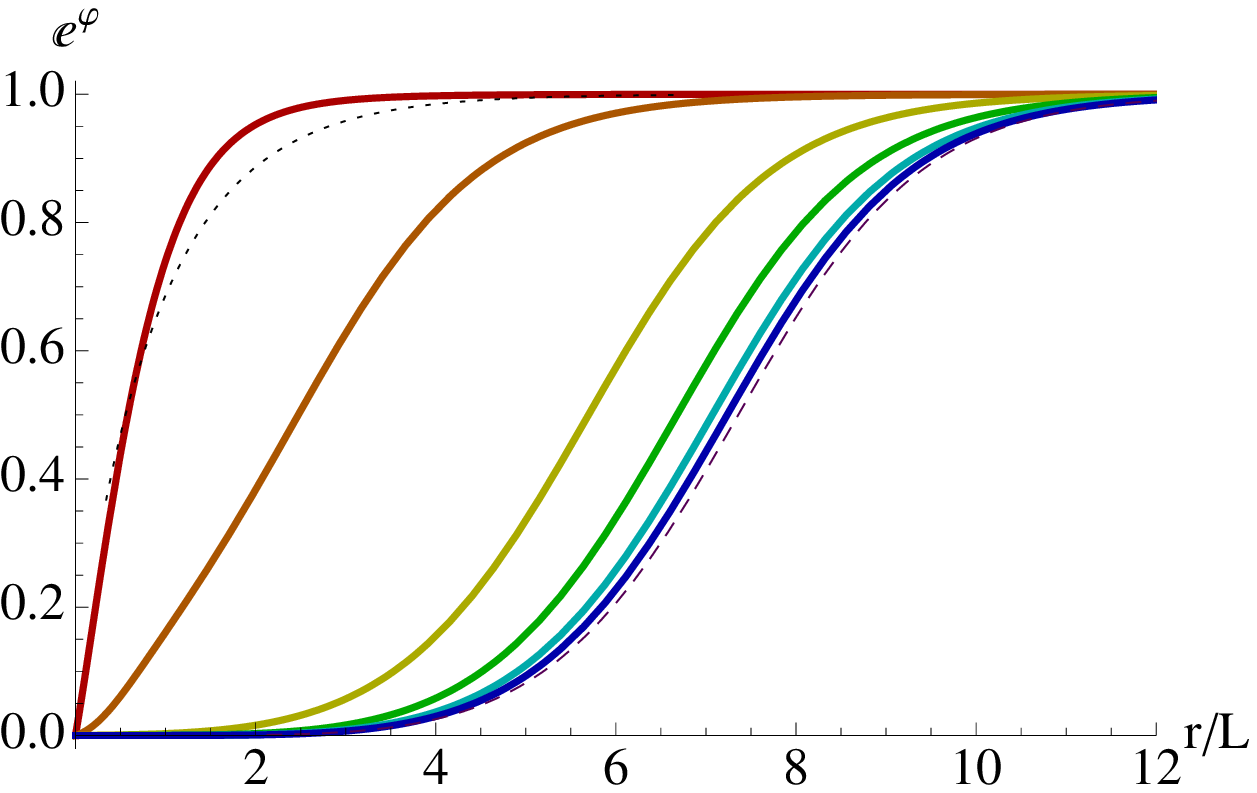}
\caption{Radial profile of the scalar in the $n=1$ model (\ref{eq.onefieldmodel})  for several values of $\omega$. Left: the cases $\omega=0$ (dashed blue curve) and $\omega=\tfrac{9}{10}\tfrac{\pi}{2}$. Right: result for $\omega = \left(1-10^{-(2k+1)}\right)\tfrac{\pi}{2}$ for $k=0$ (leftmost solid line) to $k=5$ (rightmost solid line). The dotted (dashed) line in the left (right) corresponds to the $\omega =0$ $n=1$ ($n=7$) profile in figure \ref{fig.scalars}. }\label{fig.nis1case}
\end{center}
\end{figure}

We used this prescription to determine the value of $\beta_2$, at finite $\omega$, for which the second order differential equations give rise to the supersymmetric solution. As above, the latter is given by the limiting value between the regions in which $\varphi(r)=0$ has just one asymptotic solution or two: one asymptotic and one at finite $r$. In figure  \ref{fig.nis1case}, a plot of the radial profile of the scalar in the supersymmetric flow is presented for various values of $\omega$. When $\omega$ is small, the change in the profile is very mild, and can be seen as a distortion of the $\omega=0$  $n=1$ flow. Only when $\omega$ gets very close to $\omega=\tfrac{\pi}{2}$, a rapid change in the profile occurs. 
We observe in figure \ref{fig.nis1case} (right) that the kink in the radial profile occurs at larger values of the radial coordinate. When $\omega=(1-10^{-11})\tfrac{\pi}{2}$, the profile is very close to the profile for $\omega=0$ and $n=7$. This is a consequence of the symmetry \eqref{eq.varphisymmetry}, and can be seen as a successful test  for our prescription for $\beta_2$. For this extreme value of $\omega$ we have to set $\beta_2=10.5$.

A change of phase from $\omega=0$ to $\omega=\tfrac{\pi}{2}$ interchanges two SO(7)-symmetric theories describing different physics, as can be seen from the study of the spectrum of a marginal scalar operator or the Wilson loops (see the appendix).
This change occurs  at the extreme values of $\omega$, and not at some intermediate value.
To see this, we calculated the potential in the Klein-Gordon equation for a probe massless scalar on all the background flow solutions presented in figure \ref{fig.nis1case}. In all these cases, the potential still implies a discrete spectrum for the marginal scalar operator, as in the case for $\omega=0$, from where we conclude that the change does indeed occur at $\omega=\tfrac{\pi}{2}$. Here, we recover the $\omega= 0$ $n=7$ solution, which leads to continuous spectrum with no mass gap.


\section{Conclusions} \label{sec.conclusions}

We have studied the behaviour under renormalisation group of the family of field theories, defined at least in the strict large $N$ limit, that are dual to the gauged supergravities of \cite{Dall'Agata:2012bb}. On the supergravity side, a phase $\omega$ selects a combination of electric and magnetic gauge fields of the $\omega=0$ electric frame of \cite{deWit1} to gauge SO(8) dyonically. New couplings in the supergravity are generated by the presence of $\omega$. In particular, the scalar potential changes with respect to the purely electric, $\omega=0$, case, and develops new AdS extrema. The large $N$ field theories associated to these supergravities accordingly develop a rich conformal phase structure. We have exhaustively characterised the RG evolution of the dual field theories between superconformal phases with at least SU(3) global invariance, triggered by supersymmetric mass deformations of the maximally supersymmetric, SO(8)-symmetric phase. The result is a rich $\omega$-dependent  pattern of unique and one-parameter families of flows interpolating between these superconformal phases. We have also initiated the study of the Coulomb branch of these large $N$ field theories.

The flows between superconformal phases display three types of patterns, depending on whether the dyonically-gauging angle attains either extremum, $\omega = 0$ or $\omega = \frac{\pi}{8}$, or strictly lies within its defining range. We have summarised this situation in table \ref{Table:cones}  in the introduction. For generic values of $\omega$, the critical points do not display symmetries in their positions in scalar space nor in the values of the cosmological constants. At both extrema of the interval for $\omega$, however, the critical points do arrange themselves in symmetric configurations in field space, in such a way that symmetry-related points turn out to have the same cosmological constants. These symmetric points thus describe dual phases with the same spectra and free energies, and are thus indistinguishable on physical grounds. This leads us to $\mathbb{Z}_2$-identify symmetry-related critical points and their associated flows for extreme values of $\omega$. 

At $\omega=0$, we recover the pattern of flows studied in \cite{Bobev:2009ms}. It is interesting to observe that, in this case, these $\mathbb{Z}_2$ identifications can indeed be explicitly understood from the field theory. As argued in \cite{Bobev:2009ms} from a field theory analysis in terms of BLG theory \cite{Bagger:2006sk}, the ($\omega=0$) cone of flows is due to a deformation of the BLG, $\cN=8$, SO(8) symmetric phase by two $\cN=1$ mass terms,
 \begin{equation} \label{eq:BLGsketch}
 S_{SO(8)} \to S_{SO(8)} + \frac{1}{2} m_1 {\cal O}_1 +  \frac{1}{2} m_2 {\cal O}_2 \ ,
 \end{equation}
 that, accordingly, break supersymmetry down to $\cN=1$ generically and trigger an RG flow. The parameters $m_1$ and $m_2$ qualitatively correspond to our $\lambda_-$ and $\lambda_+$ in section \ref{w=0Flows}. When one of the mass parameters, $m_1$ or $m_2$ are set to zero, the flow proceeds in a G$_2$, $\cN=1$ invariant manner to a G$_2$ phase, described in the supergravity by either of the two geometric G$_2$ critical points. These must be identified, since which one is reached depends on whether $m_1=0$ or $m_2=0$, and this is just a relabelling of $m_1$ and $m_2$. When both parameters are set equal, $m_1 = m_2$, the flow proceeds in an SU(3)$\times$U(1) invariant way towards the $\cN=2$ point with that symmetry. The U(1) factor, corresponding to the R-symmetry of the dual field theory with enhanced $\cN=2$, now rotates the mass terms in (\ref{eq:BLGsketch}). Finally, when $m_1 \neq m_2$ and are both non-zero, the flow proceeds in an SU(3) $\cN=1$ invariant manner towards the SU(3)$\times$U(1) point. Any field theory interpretation of the new cones that arise for non-zero $\omega$ should likewise take into account the lack, when $0<\omega<\frac{\pi}{8}$, or again enhanced, when $\omega= \frac{\pi}{8}$, symmetry structure of the dual phases. Similarly, the $\omega$-dependent boundaries of the new cones should be properly implemented in the dual field theory. 
 
Uplifting formulae of the original $\cN=8$ supergravity \cite{deWit1} into eleven dimensions permit the discussion of the $\omega =0$ flows from an M-theory and fully-fledged dual field theory perspectives. The  $\omega =0$ domain walls between AdS critical points of section \ref{sec.flows} uplift to $D=11$ solutions interpolating between two AdS M-theory solutions. The $\omega =0$ singular flows of section \ref{sec.coulombbranch} that run off to infinite values of the scalar fields instead uplift to configurations, regular from a $D=11$ perspective, of continuously distributed M2-branes which are, thus, dual to Higgsed phases of the field theory. It is  tempting to borrow this intuition to speculate that evolution in $\omega$ should turn some of the $\omega =0$ Coulomb flows into flows between distinct superconformal phases of the large $N$, $\omega \neq 0$ dual field theories. This indeed is the picture that emerges from our results of section \ref{sec.flows}.  The three cones of flows between critical points  that we find at finite, generic $\omega$, collapse as $\omega \rightarrow 0$, leaving just a single cone. In the $\omega \rightarrow 0$ limit, the IR endpoints of the new, $\omega \neq 0$ cones are pushed all the way out to infinity (or the boundary of the disks in the parametrisation of section \ref{sec.flows}), and so are the flows they dominate in the IR. A transition thus occurs at $\omega =0$, turning $\omega >0$ flows between superconformal points into $\omega =0$ Coulomb flows. This transition, that would render AdS-interpolating solutions into  smeared brane solutions, should be also interesting to investigate from a higher-dimensional perspective if continuous uplifiting formulae for $\omega \neq 0$ were found.

It is also interesting to observe the behaviour of the cosmological constant along generic flows inside given cones, plotted in figure \ref{fig.potentialpiover8}. The cosmological constant as a function of the radial coordinate along a flow develops increasingly longer plateaux, dominated by the values of the cosmological constants at limiting G$_2$ points, for flows with trajectories increasingly distant from the direct flow and closer to the G$_2$ points. These plateaux are a required ingredient in, for example, holographic models of walking technicolour (see  \cite{Piai:2010ma} for a review). Phenomenology, however, requires a confining phase in the IR for these models, which our CFT-interpolating flows of section \ref{sec.flows} obviously lack. Interestingly, the second-order, non-supersymmetric Coulomb branch flows of the $n=1$ model considered in section \ref{sec.coulombbranch} also display similar plateaux dominated by the non-supersymmetric SO(7) point. Although these models run to infinite values of the scalar fields, they do not lead the dual theory into confining phases either. 

It would be worth determining  whether the (large $N$) field theories dual to the family of supergravities of \cite{Dall'Agata:2012bb} display these interesting phases, or if slightly modified supergravity models do. It would be also interesting to extend our analysis into other regimes of the dual field theories, like finite temperature and chemical potentials.


\section*{Acknowledgements}

Part of this work was carried out during the 2013 {\it Strings} and {\it Gravity} Benasque workshops. We would like to thank Carlos N\'u\~nez for useful discussions and comments on the draft, and Bernard de Wit, Daniel Jafferis, Krzysztof Pilch, Henning Samtleben, Joan Simon, Matt Strassler and Nick Warner for helpful discussions. J.T. is supported by MEC FPA2010-20807-C02-02, by ERC StG HoloLHC - 306605 and by the Juan de la Cierva program of the Spanish Ministry of Economy. O.V. is supported by a Marie Curie fellowship from the European Commission, and is grateful to the CPHT of \'Ecole Polytechnique for managing the administration of this grant.


\appendix

\section{Wilson loops in the $\omega=0$ Coulomb branch} \label{sec:Wilson}

In order to further probe the phenomenology of the different cases we are treating we will focus now on the potential between a test quark-antiquark pair, which is obtained on the gravity side with the expectation value of a Wilson loop of a string hanging from the boundary into the bulk, on a background given by any of the $\omega =0$  flows  of section \ref{sec:Coulombw=0}. We will assume the string to be extended along a coordinate, say $x^1=x^1(r)$, and at the boundary the two extrema will be separated by a distance $L_{Q \bar Q}$. The string then hangs into the bulk coordinate.

\begin{figure}[tb]
\begin{center}
\includegraphics[scale=0.38]{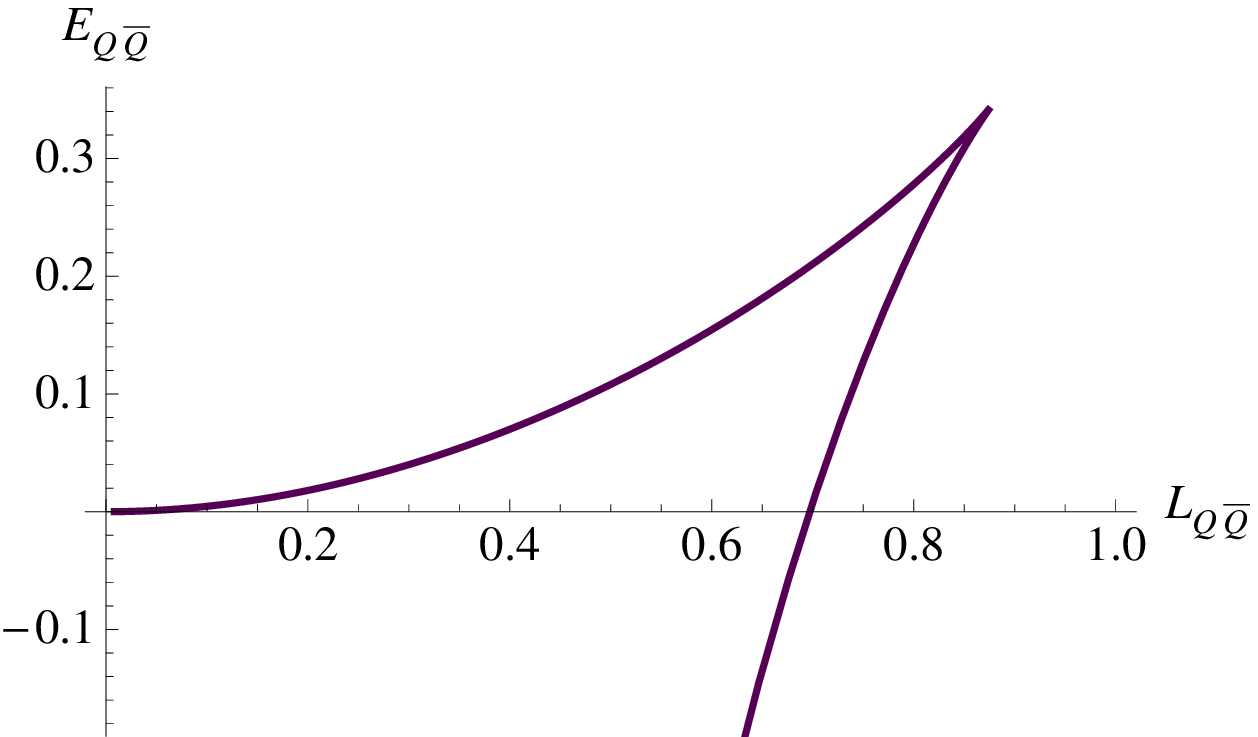}
\includegraphics[scale=0.38]{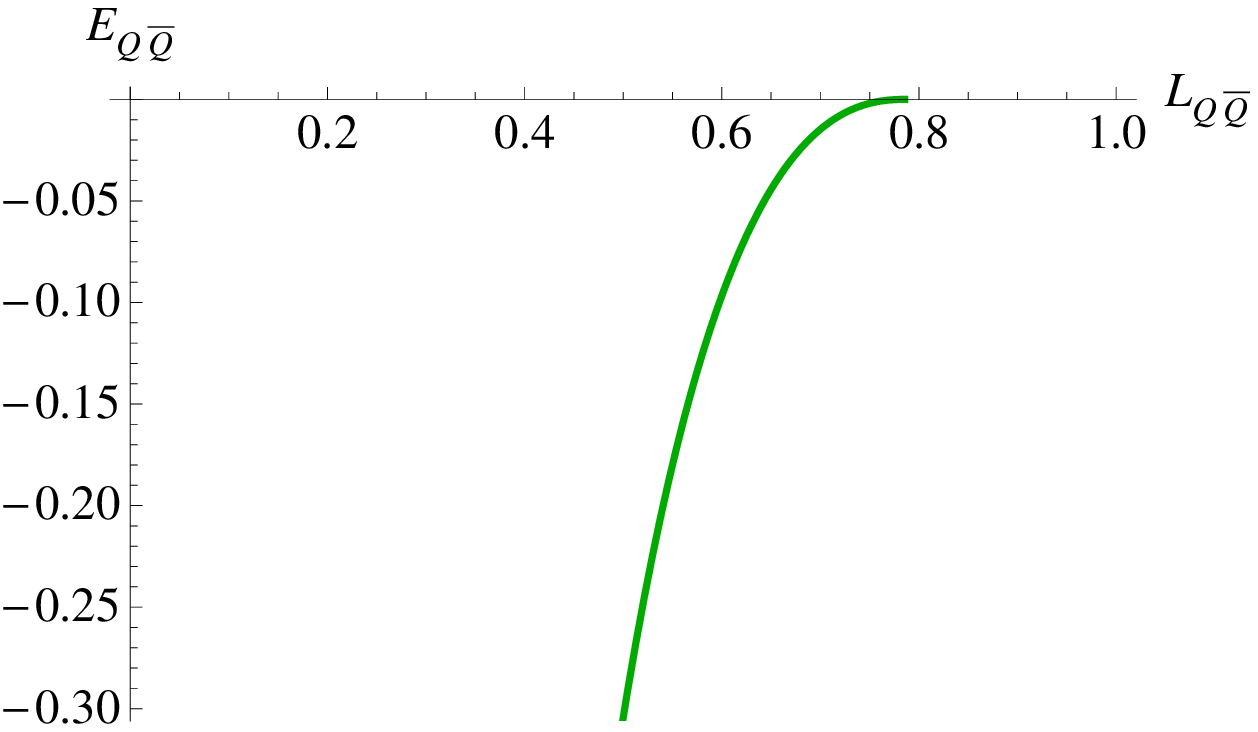}
\includegraphics[scale=0.38]{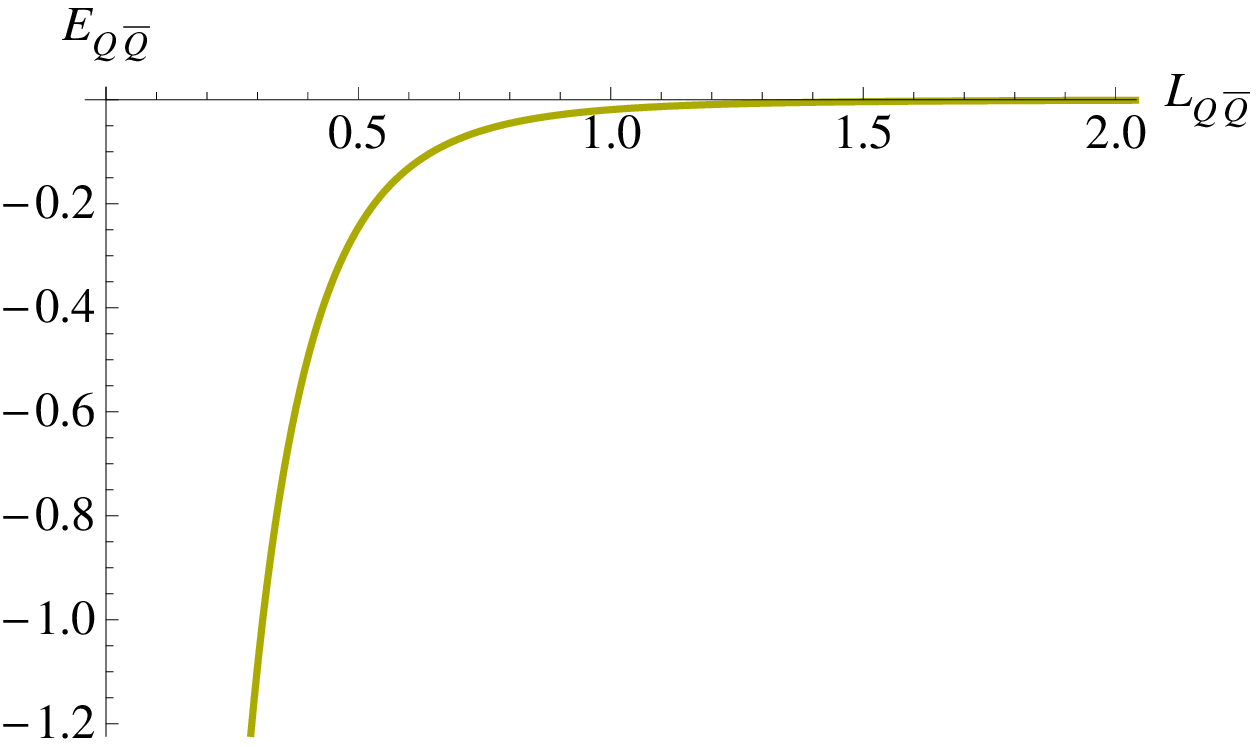}
\caption{Representative graphs of the potential between quark-antiquark as a function of their separation, for $n=1$ (left), $n=4$ (left) and $n=5$ (right).}\label{fig.wilsonloops}
\end{center}
\end{figure}

 Following the general treatment of 
\cite{Sonnenschein:1999if} we can calculate the interdistance $L_{Q \bar Q}$ and the associated energy density $E_{Q \bar Q}$ with the metric \eqref{eq:DW}. Actually, it is simpler to exploit the monotonicity of the scalar $\varphi$ to trade the $r$ integral for a integral on $\varphi$ with the aid of \eqref{eq.rvarphi}, which technically gives stability to the numeric integrations that need to be performed. In this approach the integrals are regulated by subtracting the energy of two straight strings that hang from the boundary to the IR of the geometry, with a constant separation $L_{Q \bar Q}$.
We find three different possible behaviours of the energy $E_{Q \bar Q} $ as a function of the separation length  $L_{Q \bar Q}$. For each of these we plot a representative profile in figure \ref{fig.wilsonloops}.

A common feature of all cases is the behaviour for strings that only probe the UV part of the geometry, associated to short interquark distances $L_{Q \bar Q}$. Then the potential between quarks is governed by the AdS$_4$ result, and in particular
\begin{equation}
E_{Q \bar Q} \sim \frac{1}{ L_{Q \bar Q}} \ .
\end{equation}
As in the case of the spectrum of the marginal scalar operator, it is the IR that differentiates between the different cases:
\begin{enumerate}
\item For $n<4$ we have the behaviour in the left-most graph in figure \ref{fig.wilsonloops}. We observe that for small $L_{Q \bar Q}$ there are two possible settings: a short string probing just the UV part of the geometry and a long string that extends deep into the bulk. The latter setting has positive $E_{Q \bar Q}$ and is energetically disfavoured. These two branches meet at a certain value $L_{Q \bar Q}=L_*$, but before reaching that separation the energy of the quark-antiquark pair reaches zero, at $L_{Q \bar Q}=L_c < L_*$. For $L_{Q \bar Q}>L_c$ the configuration with two straight strings hanging from the UV into the IR is favoured, which is interpreted as the signal of screening of the quark charges.
\item In the $n>4$ case, represented in the right-most graph of figure \ref{fig.wilsonloops}, the relation between $E_{Q \bar Q}$ and $L_{Q \bar Q}$ is bijective. The energy is a monotonically increasing function of $L_{Q \bar Q}$ that approaches zero as a power when the interquark distance is very large. The power is $n$-dependent, and given by
\begin{equation}
E_{Q \bar Q} \sim L_{Q \bar Q}^\frac{4}{4-n} \quad \textrm{for $n>4$ and large $L_{Q \bar Q}$.}
\end{equation}
\item When $n=4$ the relation between the interquark potential and distance is also bijective. However, now a string that hangs all the way down to the IR of the theory and comes back to the UV will have a finite distance between its extrema. Configurations corresponding to hanging strings are depicted by the thick line in the center graph of figure \ref{fig.wilsonloops}. We observe that this thick line ends at a certain $L_{Q \bar Q}=L_c$. For  interquark distances larger than this critical one the only existent configuration is that of two straight strings hanging independently. Therefore this case  represents perfect screening of the quarks, but in a different manner to the $n<4$ situation.
\end{enumerate}


\bibliography{flows}
\end{document}